\documentclass[%
 reprint,
 amsmath,amssymb,
 aps,
]{revtex4-2}

\pdfoutput=1

\usepackage{graphicx}
\usepackage{dcolumn}
\usepackage{bm}
\usepackage{lipsum} 

\graphicspath{{Figures/}}
\newcommand{\MM}{\mathbb{M}}

\begin{document}

\preprint{APS/123-QED}

\title{Dynamics of groups of magnetically driven artificial microswimmers}

\author{Jake Buzhardt}
\email{jbuzhar@g.clemson.edu}
\author{Phanindra Tallapragada}%
 \email{ptallap@clemson.edu}
\affiliation{%
 Department of Mechanical Engineering\\
 Clemson University, Clemson, SC 29630
}%




\date{\today}

\begin{abstract}
 Magnetically driven artificial microswimmers have the potential to revolutionize many biomedical technologies, such as minimally-invasive microsurgery, micro-particle manipulation, and localized drug delivery.  
 However, many of these applications will require the controlled dynamics of teams of these micro-robots with minimal feedback.  
 In this work, we study the motion and fluid dynamics produced by groups of artificial microswimmers driven by a torque induced through a uniform, rotating magnetic field. 
 Through Stokesian dynamics simulations, we show that the swimmer motion produces a rotational velocity field in the plane orthogonal to the direction of the magnetic field's rotation, which causes two interacting swimmers to move in circular trajectories in this plane around a common center. 
 The resulting over all motion is on a helical trajectory for the swimmers.  
 We compare the highly rotational velocity field of the fluid to the velocity field generated by a rotlet, the point-torque singularity of Stokes flows, showing that this is a reasonable approximation on the time average.  
 Finally, we study the motion of larger groups of swimmers and show that these groups tend to move coherently, especially when swimmer magnetizations are uniform. 
 This coherence is achieved because the group center remains almost constant in the plane orthogonal to the net motion of the swimmers.
 The results in the paper will prove useful for controlling the ensemble dynamics of small collections of magnetic swimmers.
\end{abstract}

\maketitle

\section{Introduction}


In recent years, artificial microswimmers have received significant experimental and theoretical research attention due to their promising potential biomedical applications, such as targeted drug delivery and particle manipulation \cite{nelson_arbe_2010,sitti_IEEE_2015}.  
While various means of propulsion have been considered, swimming bodies driven by externally applied magnetic fields seem particularly promising, as they present the capability of controlling the swimmers remotely \cite{nelson_cts_2017}.  
These robots maneuver through fluids at extremely small length and velocity scales, at which the effects of viscous dissipation are dominant and inertial effects are negligible.  
In this setting, the typical means of large scale propulsion are ineffective and propulsion mechanisms must instead rely on non-reciprocal gaits \cite{purcell_1977}.  
As an example seen in nature, many bacteria achieve self-propulsion using a rotating flagella or flagellar bundle in the form of a helix \cite{berg_nature_1973,Berg_AnnRev2003}.  
Taking this as inspiration, many helically-structured artificial swimmers have been designed to rotate due to a uniform, rotating magnetic field and achieve propulsion in a similar manner \cite{nelson_apl_2009,GhoshFischer_Nano2009,peyer_nanoscale_2013}. 
Further, it has been shown that this motion is more efficient than simply towing an object by a magnetic field gradient \cite{nelson_ijrr_2009}.

\begin{figure}[t]
    \centering
    \includegraphics[width = 0.49\linewidth]{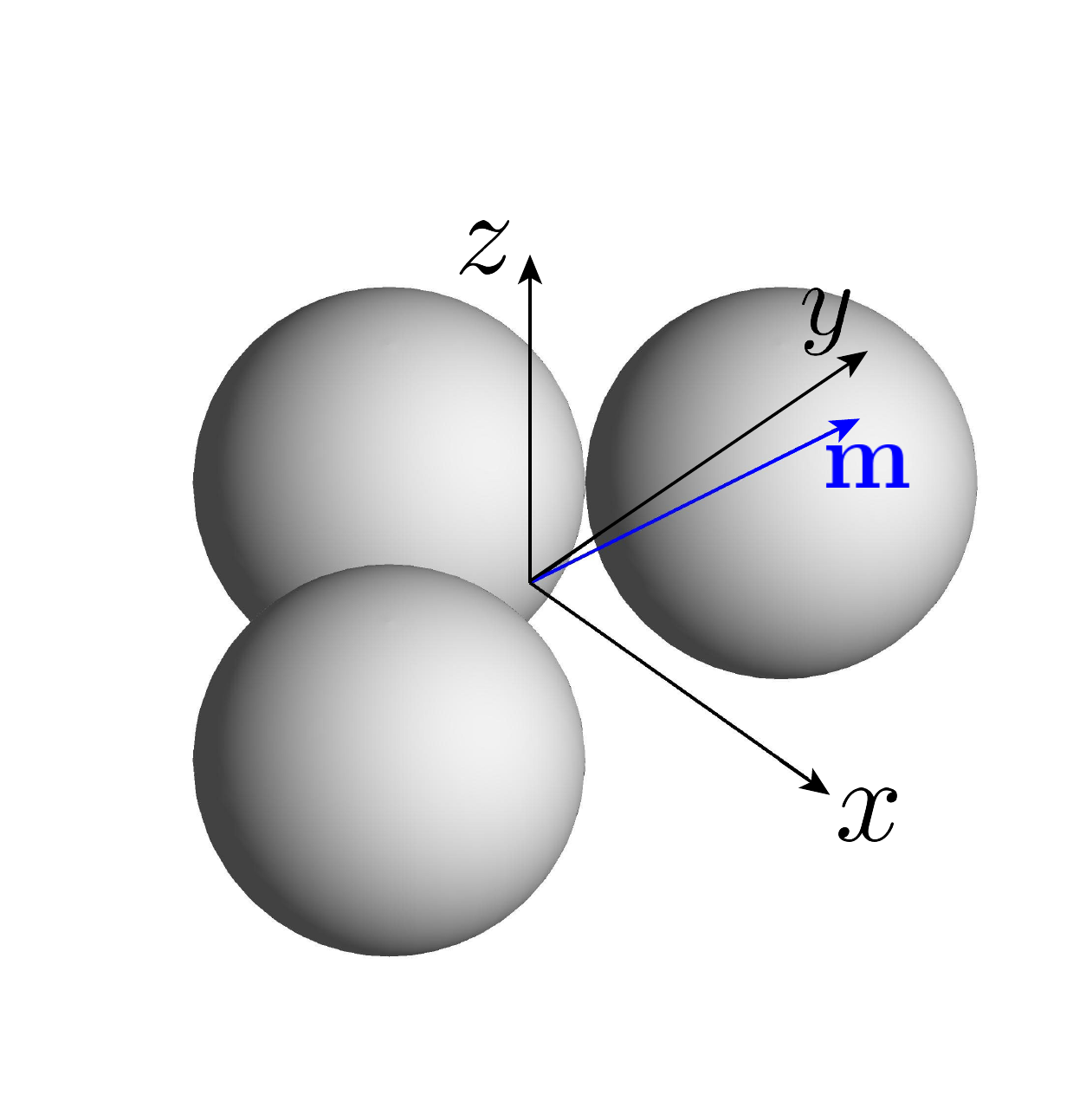}\\[0.5ex]
    (a)\\
    \includegraphics[width=0.49\linewidth]{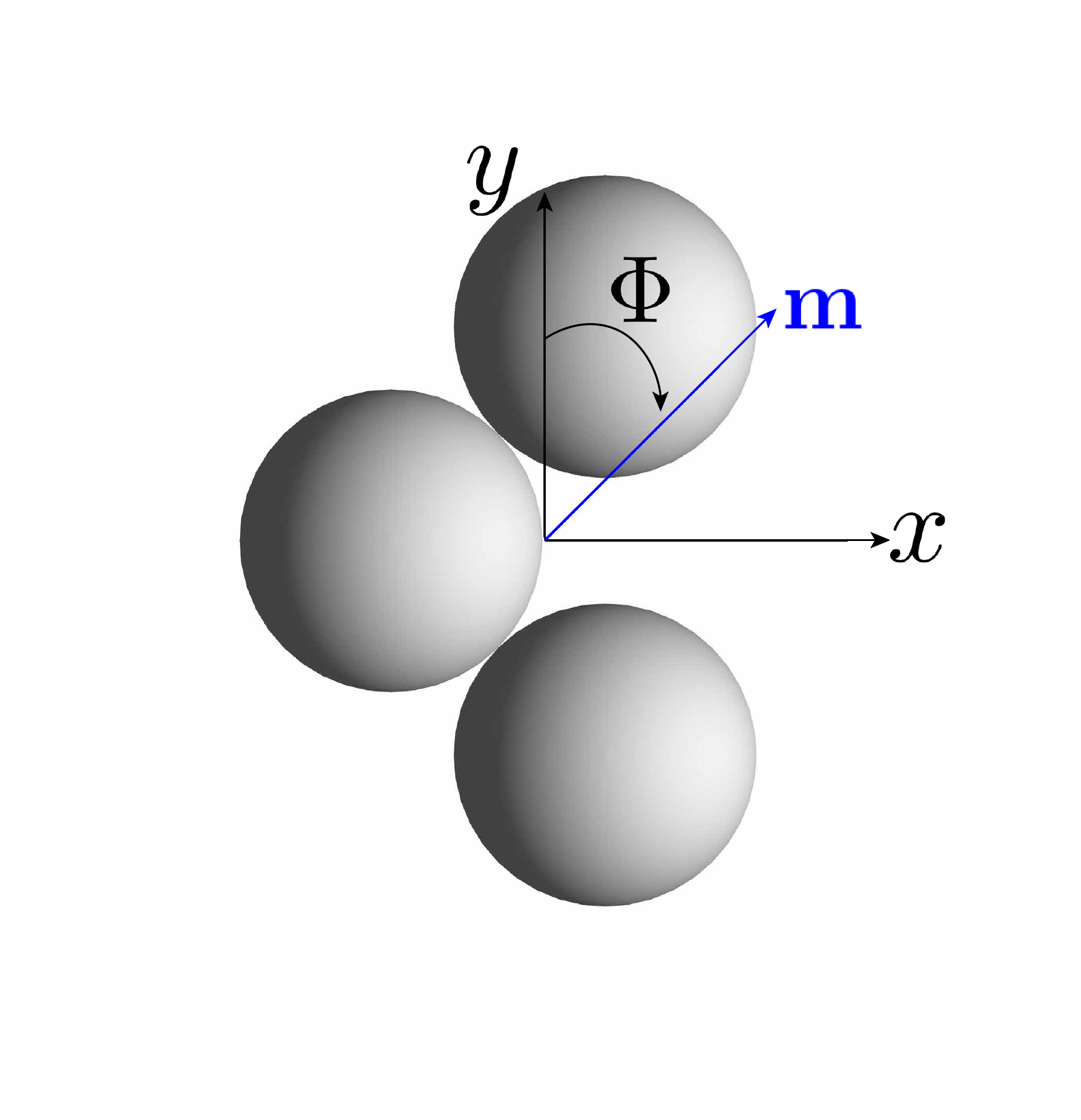}
    \includegraphics[width=0.49\linewidth]{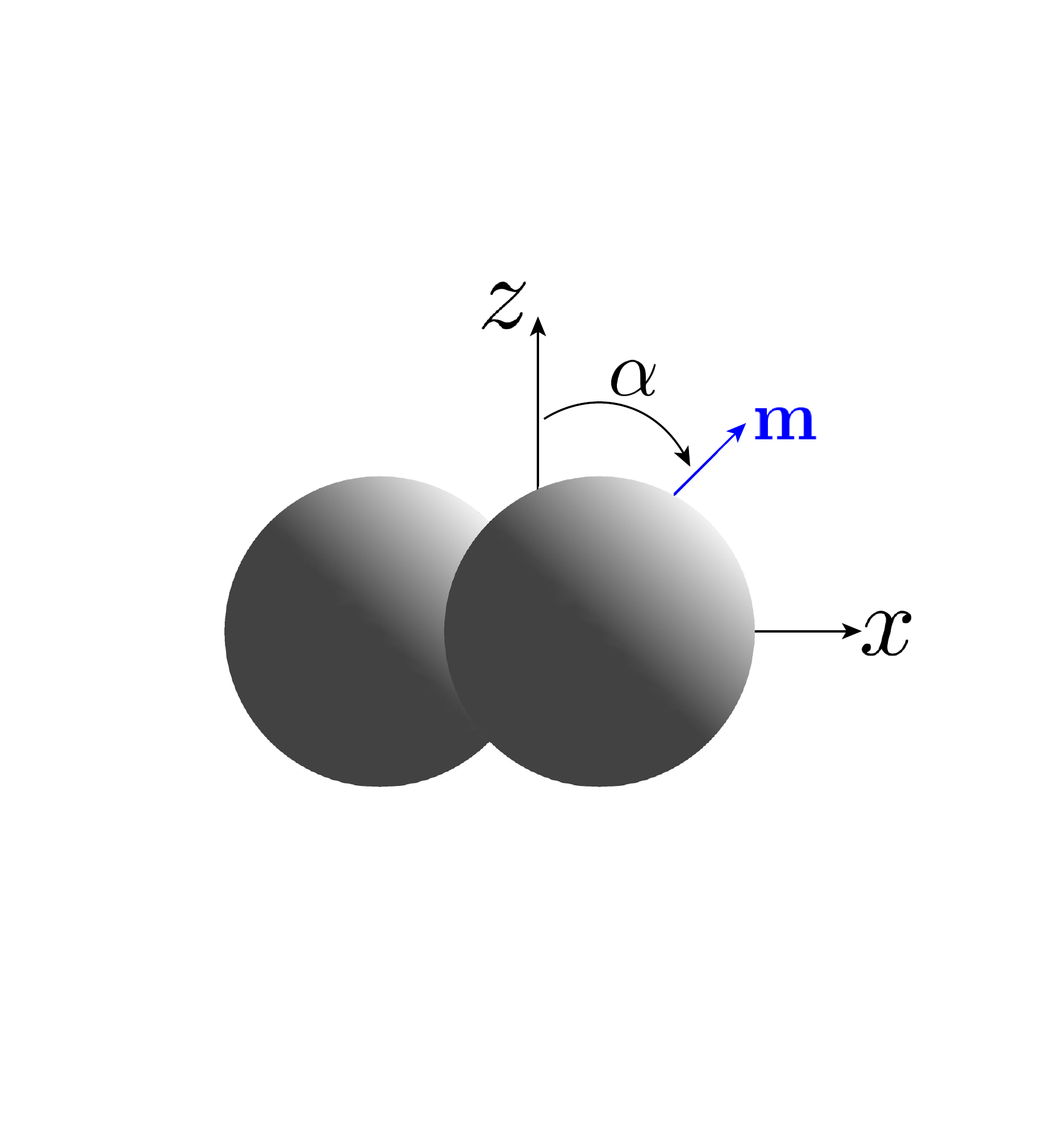}\\
    (b) \hspace{3.5cm} (c)
    \caption{Three views of the three-sphere magnetic microswimmer, shown in the body-fixed frame of reference.  $\mathbf{m}$ is the magnetic moment.  $\alpha$ and $\Phi$ are the azimuthal and polar angles used to parameterize the magnetic moment orientation.}
    \label{fig:SwimmerSchematic}
\end{figure}

Other experimental work has developed more simplistic swimmer geometries, such as groups of magnetic beads \cite{TGPS2008PRL,kim_pre_2014,Kim_SRep2016,Sing_Walkers_PNAS2010,morimoto_walkers_PRE2008}.
These micro-robots are particularly relevant due to the ease with which they may be manufactured, but also because the symmetric nature of the sphere particularly lends it to theoretical and computational modeling.  
In this work, an artificial swimmer body composed of three magnetic spheres rigidly connected in a configuration with a $90^\circ$ bend is studied; this geometry is depicted in Fig. \ref{fig:SwimmerSchematic}.  
This swimmer has been of particular interest, as it is known to be one of the simplest geometries that possesses the necessary asymmetry to couple an external torque to a translational motion \cite{happel_brenner}.  
This specific swimmer geometry was first proposed in the work of Cheang \emph{et al.} \cite{kim_pre_2014}, where fabrication and simple motion control were experimentally demonstrated. 
Meshkati and Fu \cite{fu_pre_2014} later computed the mobility matrix for this geometry using the method of regularized Stokeslets and used this to predict swimmer orientations relative to the magnetic field and provide comparisons to the experimental work of \cite{kim_pre_2014}. 
As part of a more general study of achiral propellers, Morozov \emph{et al.} \cite{Leshansky_prf_2017} considered the effect of geometry, magnetization, and driving frequency on the propulsion of this swimmer body. 

While this particular swimmer geometry has been examined both theoretically and experimentally, almost all effort has been directed towards steering and propulsion of a single swimmer body.
However, if these micro-robots are to be used in practice for targeted drug delivery, it is likely that a large number of these bodies will be required to move as a group to deliver a payload \cite{Kim_APL2014,Nelson_AHM2012}. The requirement for a large number of micro-robots is due to the fact that it is only possible to load such small bodies with very small doses of therapeutic materials.  Even so, testing of these applications has shown that relatively small proportions of the loaded quantity actually reaches the target destination \cite{Nelson_AHM2012}.  This is largely due to material limitations and difficulties associated with group control in complex environments.  For these reasons, and since manufacturing developments have led to cost effective means of producing large numbers of these micro-robots, applications of targeted delivery using teams or swarms of drug-carrying micro-robots have been proposed \cite{sitti_IEEE_2015,Pouponneau_IJRR2009,Nelson_AHM2012,Palberg_Langmuir2013,Kim_APL2014,HuangNelson_SM2014,Lanauze_IJRR2014,ServantNelson_AdvMat2015,Wang_ADFM2018}. For such swarms of these robots to be effectively used, it will be necessary to develop a more thorough understanding of the hydrodynamic effects that arise when these bodies move in groups.  

Regarding the geometry composed of three spheres, in experimental work, Cheang \emph{et al.} \cite{Kim_APL2014} demonstrated control of two geometrically similar but magnetically different swimmers using a single, uniform, rotating magnetic field.  Later, Cheang \emph{et al.} \cite{Kim_SRep2016} showed that such magnetic swimmers composed of beads can be deployed as modular sub-units, assembling and disassembling due to magnetic interactions.  
Recently the authors showed in simulation that the three-sphere artificial swimmer can be used to manipulate a passive spherical particle using the velocity field generated by the swimmer \cite{bt_micromanipulation_DSCC19}. Other experimental works have also demonstrated controlled motion of bodies composed of just two spheres, but these works typically rely on some alternative means of symmetry-breaking, such as motion next to a wall or substrate \cite{tierno_PRL2008,morimoto_walkers_PRE2008}, relative motion between two particles \cite{Biswal_SM2018}, or compositional variations in the body \cite{Ni_SM2017}.

In theoretical work, while little prior works exist regarding this particular geometry and propulsion mechanism for multiple interacting swimmers, many previous works study the motion of pairs or groups of biological \cite{ardekani_PRE2014,ishikawa_simmonds_pedley_2006,LLOPIS2010}, artificial \cite{KeavenyMaxey_PRE2010,fauci_pumpinghelices}, or hypothetical model swimmers \cite{fauci_toroid,Lauga_ManyScallop,YeomansDumbells_2008,Yeomans_PRL07_IxNajafi}.  
Using the method of regularized Stokeslets, Huang and Fauci \cite{fauci_toroid} and Buchmann \emph{et al.} \cite{fauci_pumpinghelices} studied the velocity fields produced by groups of interacting toroidal swimmers and helical mixers respectively, showing that these swimmers produce velocity fields that lead to useful fluid motion and cooperative motion of the swimming bodies.
Similarly, in this paper, Stokesian dynamics simulations are used to analyze the fluid velocity field produced by groups of magnetically-driven swimmers and show how the resulting hydrodynamic interactions lead to group and pairwise motions that differ from the case of an individual swimmer. 

Many of the works mentioned above consider either only a very small number (three or fewer) of these bodies.  
Further, most of these works \cite{ardekani_PRE2014,ishikawa_simmonds_pedley_2006,LLOPIS2010,fauci_toroid,fauci_pumpinghelices,Lauga_ManyScallop,YeomansDumbells_2008,Yeomans_PRL07_IxNajafi} prescribe a velocity, either of the body or on the surface, as a means of propulsion. 
Keaveny and Maxey \cite{KeavenyMaxey_PRE2010} implemented the Force Coupling Method along with a detailed magnetic model to study the cooperative motion of stacked and side-by-side pairs of swimmers propelled by the magnetically driven undulations of a helical flagella. 
Similarly, in this work, a detailed model of the propulsion mechanism of the magnetically driven microswimmer is used by considering a magnetic dipole that remains fixed in the co-rotating frame of reference. 
With this, a comparable simulation technique is implemented to study the motion of a different geometry in similar pair configurations and then extended to study larger groups of swimmers. 
The numerical method used here is based on the well-known Stokesian dynamics algorithm \cite{durlofsky_brady_bossis_1987, brady_afm_1988} for simulating the motion of interacting spherical particles in a Stokes flow.  
Specifically, simulations here rely on the extended version of this algorithm given by Swan \emph{et al.} \cite{SwanBrady_Teaching}, that allows for the simulation of bodies composed of spheres. 
This work also incorporates pairwise magnetic interactions and shows that the effect of these magnetic interactions is negligible compared to the dominant hydrodynamic interaction.

The simulations presented herein demonstrate that teams of magnetic swimmers propelled by periodic magnetic fields tend to move in a coherent fashion with the group center of the collection being nearly fixed in the plane orthogonal to the net motion with the geometric distribution of the swimmers not being distorted significantly. 
We show that this is the result of the hydrodynamic interaction between a pair of swimmers being very similar to the interaction of two rotlets, or point-torque singularities of the Stokes flow.
The results in this paper are significant for the development of dynamic  models for teams of robotic swimmers that need to move in a coordinated fashion with minimal feedback.

\section{Mathematical model}
\subsection{Hydrodynamic model}
The motion of the artificial swimmers considered in this work is described by very small length and velocity scales.  
In such a setting, the fluid motion generated by the particles is dominated by viscous forces, while inertial forces are considered negligible.  
Therefore, the fluid dynamics are well-described by the Stokes equations.

The linearity of the Stokes equations implies that the velocities of a body $A$ moving through a viscous fluid are linearly proportional to the forces and torques acting on the body.  
This is known as the mobility relationship and is given here by 
\begin{equation}
    \begin{pmatrix}
    \mathbf{V}_A\\
    \boldsymbol{\Omega}_A
    \end{pmatrix}
    =
    \begin{pmatrix}
    \mathbf{K}  &   \mathbf{C}\\
    \mathbf{C}^{\intercal}    &   \mathbf{M}
    \end{pmatrix}
    \begin{pmatrix}
    \mathbf{F}^e_A\\
    \mathbf{T}^e_A
    \end{pmatrix}
    \label{eq:Mobility}
\end{equation}
where $\mathbf{F}^e_A$ and $\mathbf{T}^e_A$ are the external forces and torques acting on the body $A$ and $\mathbf{V}_A$ and $\boldsymbol{\Omega}_A$ are the velocity and angular velocity of the body.  
The self-mobilities $\mathbf{K}$, $\mathbf{C}$, and $\mathbf{M}$ are $3\times3$ matrices that describe the hydrodynamic effects on the body's motion and depend solely on the geometry of the body. 
The mobility relationship for this particular geometry of three spheres has been derived previously in Refs \cite{fu_pre_2014,Leshansky_prf_2017}.  

To define the mobility matrix for the system, the commonly used Stokesian dynamics algorithm \cite{brady_afm_1988,durlofsky_brady_bossis_1987,SwanBrady_Teaching} is implemented. 
This algorithm was derived specifically to model the motion and hydrodynamic interaction of systems of spheres moving in a Stokes flow.  
Full details of this method are given in \cite{brady_afm_1988,durlofsky_brady_bossis_1987,SwanBrady_Teaching}, but a brief overview is given here. 
This method considers the disturbance fluid velocity field $\mathbf{u}(\mathbf{x})$ produced by each sphere in the simulation as a multipole expansion about the center of the sphere
\begin{equation}
\begin{split}
    u_i(\mathbf{x}) = \frac{1}{8\pi\mu}\Bigg[ \bigg(1+&\frac{a^2}{6}\nabla^2 \bigg)J_{ij} \, F_j + R_{ij}\, T_j \\
    &+ \left(1+\frac{a^2}{10}\nabla^2 \right)K_{ijk}\, S_{jk} + \cdots \Bigg]
    \end{split}
    \label{eq:multipole}
\end{equation}
Here $J_{ij}$, $R_{ij}$ and $K_{ijk}$ are the propagators associated with the singularities of Stokes flow, the Stokeslet, rotlet, and stresslet respectively \cite{brady_afm_1988}. The coefficients $F_j$, $T_j$, and $S_{jk}$ correspond to components of the force, torque, and stress acting on each sphere.  
The ellipsis in Eq. \eqref{eq:multipole} indicates that an infinite series of force moments would be required to exactly represent this disturbance velocity field \cite{kim_karrila}.  Following Durlofsky \emph{et al.} \cite{durlofsky_brady_bossis_1987}, in this work, the multipole expansion is truncated at the order shown in Eq. \eqref{eq:multipole}.
This expression is then used in conjunction with the Fax\'en formulae for spheres to develop the terms of a many-sphere mobility matrix.  This matrix couples the forces, torques, and stresses acting on the spheres to the velocities and angular velocities of the spheres and the rate of strain of the flow. 
Full details of this calculation, along with expressions for the elements of this many-sphere grand mobility matrix are given by Durlofsky \emph{et al.} \cite{durlofsky_brady_bossis_1987}.

While the Stokesian dynamics algorithm presented by Durlofsky \emph{et al.}\,\cite{durlofsky_brady_bossis_1987} gives the components of the many-sphere grand mobility matrix, we are interested in the motion of rigid bodies composed of spheres.  
To condense the many-sphere mobility matrix into a mobility matrix of the form in Eq. \eqref{eq:Mobility}, the constraints of rigid body motion through a Stokes flow must be applied.  
Specifically, it is required that the velocities of each sphere (denoted by $\alpha$) are related to the velocities of the rigid body (denoted by $A$) by 
\begin{subequations}
\begin{align}
\mathbf{V}_{\alpha} &= \mathbf{V}_A + \boldsymbol{\Omega}_A\times \mathbf{r}_{A\alpha}\\
\boldsymbol{\Omega}_{\alpha} &= \boldsymbol{\Omega}_A.
\end{align}
\label{eq:RigidVel}
\end{subequations}
Also, it is required that the forces and torques acting on the body satisfy the quasi-static conditions of motion in a Stokes flow 
\begin{subequations}
\begin{align}
\mathbf{F}^e_A &- \sum_{\alpha\in A}\mathbf{F}_{\alpha} = \mathbf{0}\\
\mathbf{T}^e_A &- \sum_{\alpha\in A}\bigg( \mathbf{r}_{A \alpha} \times \mathbf{F}_{\alpha} + \mathbf{T}_{\alpha}\bigg) = \mathbf{0}.
\end{align}
\label{eq:QuasiStatic}
\end{subequations}
In Eqs. \eqref{eq:RigidVel} and \eqref{eq:QuasiStatic}, $\mathbf{F}_{\alpha}$ and $\mathbf{T}_{\alpha}$ are the forces and torques acting on each sphere $\alpha$, and $\mathbf{r}_{A\alpha}$ is the vector from a chosen reference point on the rigid body $A$ to the center of the sphere $\alpha$.  
For the computations presented herein, the center of hydrodynamic mobility, as defined by Morozov \emph{et al.} \cite{Leshansky_prf_2017} is chosen as the reference point. 

\subsection{Rigid body dynamics}
To define the rotation transformation between the body-fixed frame and the spatially-fixed frame, the orientation of the artificial swimmer is parameterized using the $ZXZ$ Euler angles, where the transformation matrix from the spatial frame to the body frame is defined as 
\begin{equation}
    \mathbf{R} = \mathbf{R}_z(\psi)
    \mathbf{R}_x(\theta)
    \mathbf{R}_z(\phi)
\end{equation} 
where $\mathbf{R}_x$ and $\mathbf{R}_z$ are the matrices representing the rotation by the given angles about the body-fixed $x$ and $z$ axes respectively.  
Since these matrices are orthogonal, the inverse transformation from the spatial frame to the body-fixed frame is given by $\mathbf{R}^{-1} = \mathbf{R}^T$.

Once the angular velocities are found through the mobility relationship, they may be related to the rates of change of the Euler angles to determine how the orientation of the swimming body changes due to the magnetic field.   
This relationship is given by \cite{goldstein}
\begin{equation}
    \boldsymbol{\Omega}_A = 
    \mathbf{R}^T
    \begin{pmatrix} 
    \sin{\psi}\sin{\theta}  &   \cos{\psi}  &   0\\[0.5ex]
    \cos{\psi}\sin{\theta}  &   -\sin{\psi}  &   0\\[0.5ex]
    \cos{\theta}    &   0   &   1
    \end{pmatrix}
    \begin{pmatrix}
    \dot{\phi}\\[0.5ex]
    \dot{\theta}\\[0.5ex]
    \dot{\psi}
    \end{pmatrix}
    \label{eq:AngleRates}
\end{equation}
where the dot notation indicates the time rate of change.

This formulation allows the equations governing the motion of the swimmer to be integrated numerically by forming the mobility matrix using the Stokesian dynamics method at each timestep, computing the torque applied to the body, and using these to determine the velocity and the orientation rate of change of the swimmer.

\subsection{Magnetic actuation}
The propulsion mechanism for the artificial swimmers considered here is an applied torque $\mathbf{T}^e$ generated by a magnetic field $\mathbf{B}$.  
With the magnetic moment vector $\mathbf{m}$ defined in the body-fixed frame of reference and assumed to be permanent, the torque acting on the body is defined in the spatial frame as 
\begin{equation}
    \mathbf{T}^e = \mathbf{R}^T\mathbf{m}\times \mathbf{B}
    \label{eq:Torque}
\end{equation}
where $\mathbf{B}$ is the magnetic field defined in the spatially fixed frame. 

Here, a uniform magnetic field rotating in the $xy$-plane about the spatially-fixed $z$-axis is considered as
\begin{equation}
    \mathbf{B}(t) = B\cdot (\cos\omega t ,\: \sin\omega t,\: 0)^T
    \label{eq:MagField}
\end{equation}
where $B$ is the strength of the magnetic field and $\omega$ is the frequency of rotation. With this, the torque acting on the swimmer at any time may be calculated and used in conjunction with the mobility formulation in Eq. \eqref{eq:Mobility} to determine the velocity of the swimmer.  
In general, when multiple magnetic dipoles are present a force will arise as a result.  
In this work, such magnetic interactions are neglected, as the resulting velocities are of negligible magnitude in comparison to those resulting from hydrodynamic interaction, as is shown in later calculations. 

\begin{figure}[]
    \begin{center}
    \includegraphics[width = \linewidth]{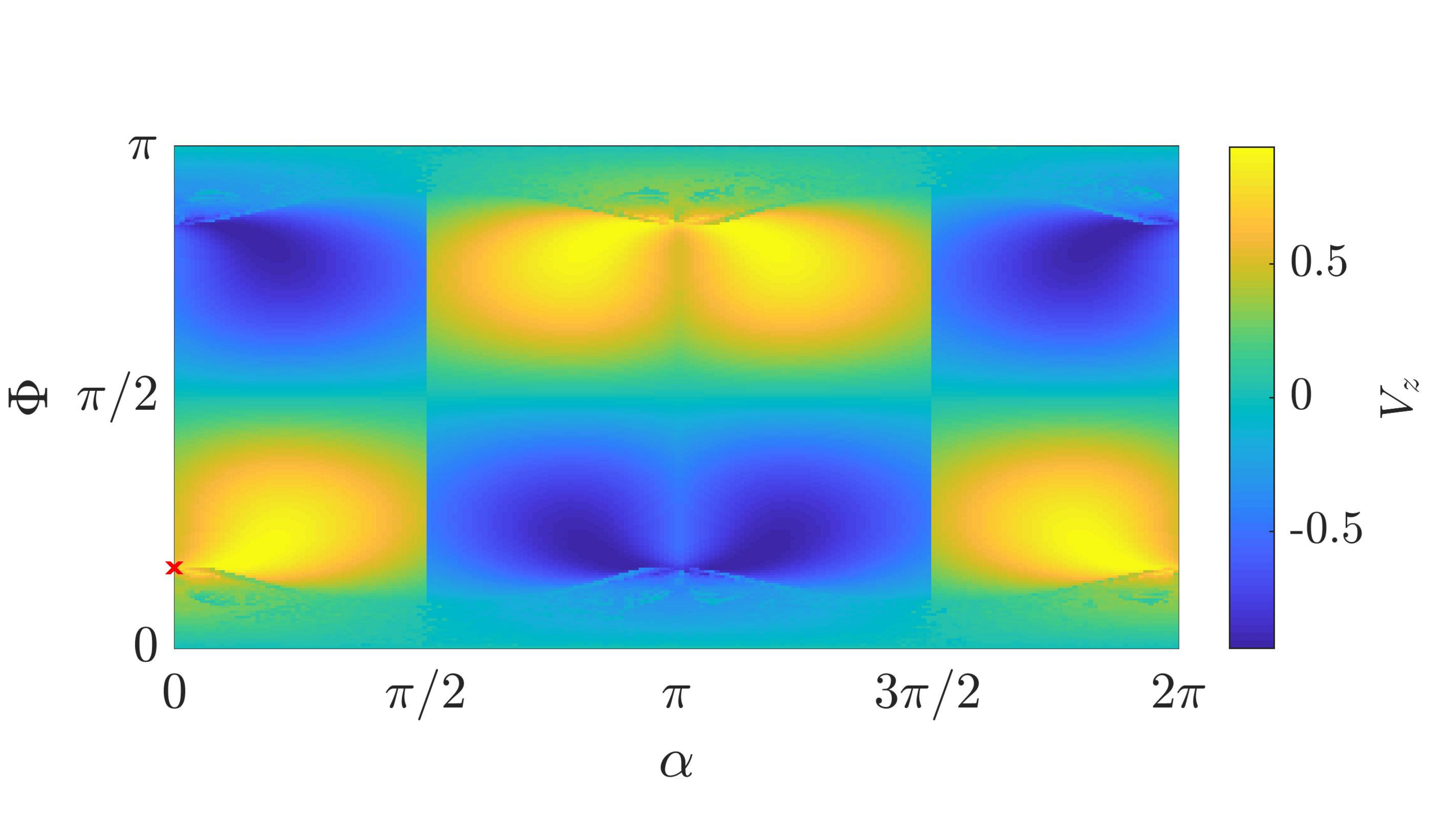}\\
    (a)\\
    \includegraphics[width=0.49\linewidth]{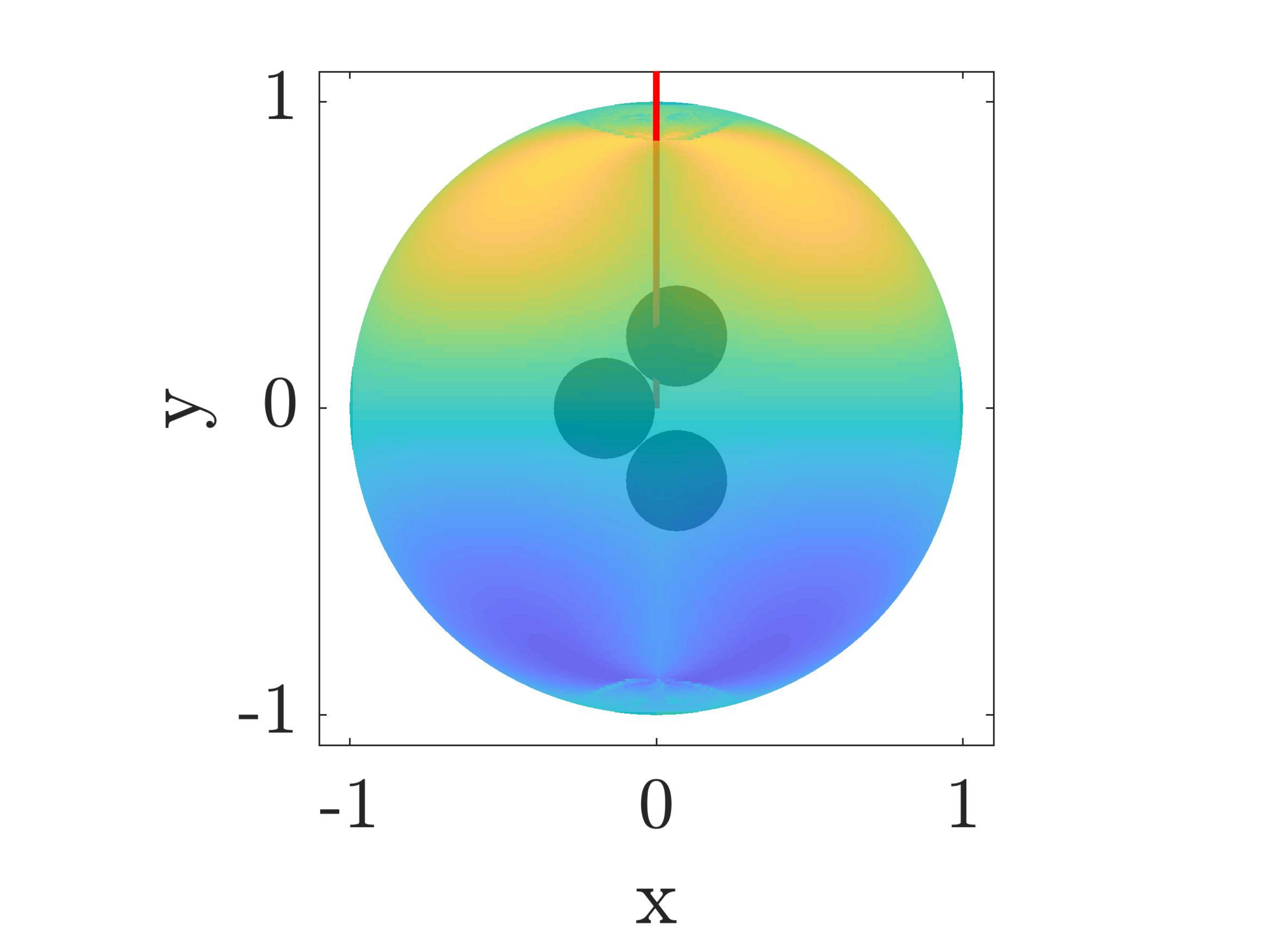}
    \includegraphics[width=0.49\linewidth]{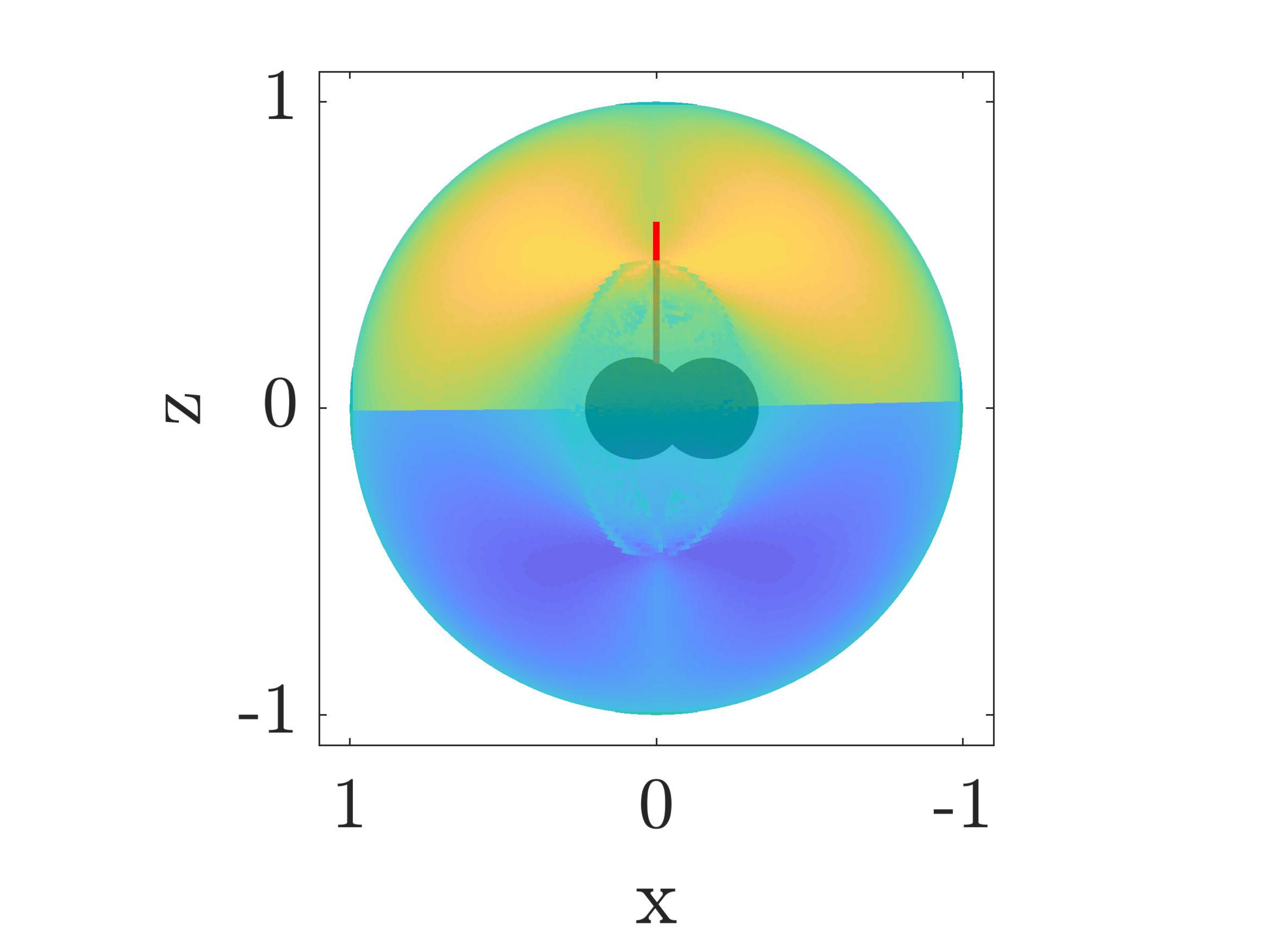}
    \end{center}
    \hspace{0.2cm}(b) \hspace{3.7cm} (c)
    \caption{Propulsion speed of the artificial swimmer as a function of the magnetic moment direction. (a) Direction parameterized by the spherical coordinates $\alpha$ and $\Phi$.  (b) and (c) show the same dependence plotted on the unit sphere.  The black $\times$ in (a) and the red line in (b) and (c) indicate the moment direction used in simulations herein. }
    \label{fig:MomentDirection}
\end{figure}

With the magnetically induced torque defined, the expressions for the spatial frame translational velocity and the orientation rate of change are fully defined.  
Integrating these equations enables the study of the effect of the magnetization of the swimmer and the applied magnetic field on the propulsion of the artificial swimmer. 
Fig. \ref{fig:MomentDirection} shows the propulsion velocity of the artificial swimmer as the direction of the magnetic moment vector is varied.  
Here the magnetic moment direction is parameterized using the azimuthal and polar angles $\alpha$ and $\Phi$ as 
\begin{equation}
    \mathbf{m} = m\:(\sin\Phi \sin\alpha ,\: \cos\Phi ,\: \sin\Phi\cos\alpha)^T
\end{equation}
For this study, the rotating magnetic field has a frequency of $\omega = 4\pi$ rad/s and strength $B = 5.0\,\text{mT}$ and the magnetic moment has a constant strength of $m = 4\times10^{-15}\,\text{J/T}$.  
Based on these simulations, a magnetic moment orientation may be chosen to obtain a desired propulsion velocity.  
Specifically, in this work, the direction given by $\alpha = 0$ rad and $\Phi = 0.5052$ rad is used for the remainder of the simulations shown, unless otherwise noted.  
This location is denoted by the red $\times$ in Fig. \ref{fig:MomentDirection} (a) and the red line in Fig. \ref{fig:MomentDirection} (b)-(c).  
It should be noted that this result depends on the driving frequency $\omega$ chosen.  
As shown by Morozov \emph{et al.} \cite{Leshansky_prf_2017}, the largest swimmer velocities will occur at the \emph{step-out} frequency for magnetization orientations with $\alpha=0$.

In Fig. \ref{fig:MomentDirection} and in the remainder of this paper, all numerical results will be reported in a nondimensional form.  
For this, the characteristic velocity $V^* = mBC_{yz}$ is adopted, where $C_{yz} = C_{zy}$ is the nonzero element of the coupling matrix $\mathbf{C}$.  
This submatrix only possesses this symmetry in the body-fixed frame due to the choice of the hydrodynamic center of mobility as the reference point.  
The characteristic length $L^*$ is chosen as the diameter of one of the magnetic spheres composing the body, $L^*=d$.  
With this, the characteristic time is chosen to be $t^* = L^*/V^*.$  
For the sake of readability, no special notation will be used for these nondimensional quantities, but all numerical values reported are normalized by these characteristic values. 

\begin{figure}[]
    \centering
    \includegraphics[width=0.9\linewidth]{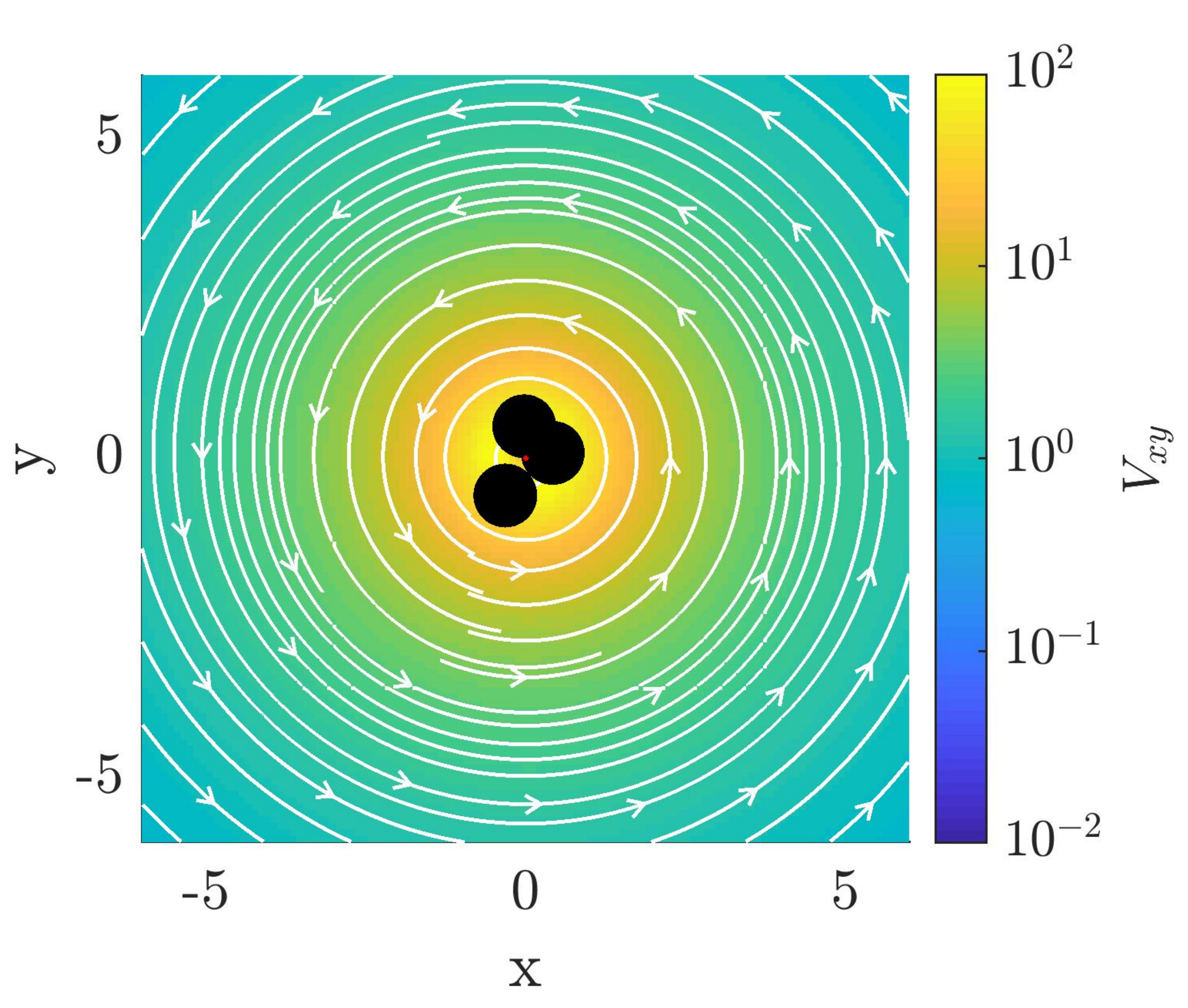}\\
    \includegraphics[width=0.9\linewidth]{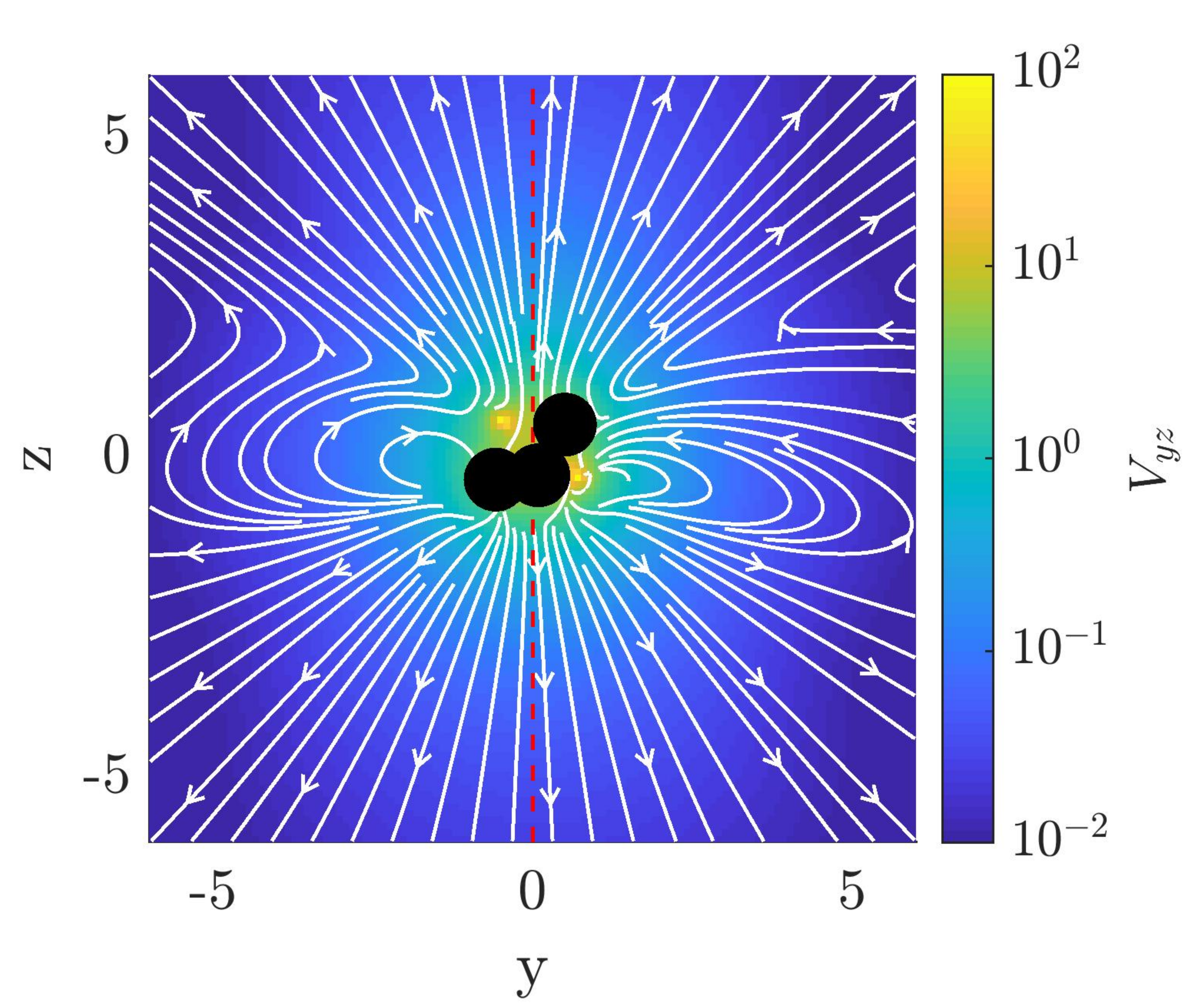}
    \caption{Streamlines and magnitude of the disturbance fluid velocity field produced by the three-sphere swimmer at steady state, time-averaged over one period of the magnetic field rotation.}
    \label{fig:VelField1}
\end{figure}

With the motion of the swimmer known from numerical integration, the fluid velocity field produced by the swimmer motion may be calculated from Eq. \eqref{eq:multipole}, where the singularity strengths $F_i$, $T_i$ and $S_{jk}$ are computed from the many-sphere grand resistance matrix.  
The fluid velocity field produced by the swimmer, time-averaged over a period of the magnetic field rotation, is shown in Fig. \ref{fig:VelField1}. 
The color represents the magnitude of the fluid velocity in the plane shown, on a logarithmic scale.  
From this figure, it can be seen that the fluid velocity decays much more rapidly in the planes containing the swimmer's axis of rotation than in the plane orthogonal to this axis. 
Also, the fluid motion in this orthogonal plane is highly rotational, indicating that the rotlet term is the dominant term in the multipole expansion of Eq. \eqref{eq:multipole}.  

\section{Multiple interacting swimmers}

A significant advantage of using the Stokesian dynamics method is that it may easily be extended to consider many interacting bodies in the low Reynolds number fluid. 
To do this, it is necessary to construct a many-body grand mobility matrix, which depends on the relative positions and orientations of the bodies.  
For the case of $n$ bodies, the matrix takes the following form.
\begin{equation}
    \begin{pmatrix}
    \mathbf{V}_1\\
    \vdots\\
    \mathbf{V}_n\\
    \boldsymbol{\Omega}_1\\
    \vdots\\
    \boldsymbol{\Omega}_n\\
    \end{pmatrix}
    =
    \begin{pmatrix}
    \mathbf{K}_{11}  & \cdots & \mathbf{K}_{1n}  & \mathbf{C}_{11}  & \cdots & \mathbf{C}_{1n} \\
    \vdots & \ddots & \vdots & \vdots & \ddots & \vdots \\
    \mathbf{K}_{n1}  & \cdots & \mathbf{K}_{nn}  & \mathbf{C}_{n1}  & \cdots & \mathbf{C}_{nn} 
    \\
    \mathbf{C}^{\intercal}_{11}  & \cdots & \mathbf{C}^{\intercal}_{1n}  & \mathbf{M}_{11}  & \cdots & \mathbf{M}_{1n} \\
    \vdots & \ddots & \vdots & \vdots & \ddots & \vdots \\
    \mathbf{C}^{\intercal}_{n1}  & \cdots & \mathbf{C}^{\intercal}_{nn}  & \mathbf{M}_{n1}  & \cdots & \mathbf{M}_{nn} \\
    \end{pmatrix}
    \begin{pmatrix}
    \mathbf{F}^e_1\\
    \vdots\\
    \mathbf{F}^e_n\\
    \boldsymbol{T}^e_1\\
    \vdots\\
    \boldsymbol{T}^e_n\\
    \end{pmatrix}
    \label{eq:GrandMobility}
\end{equation}

In simulation, this matrix is formed in a similar manner to the single swimmer mobility matrix in Eq. \eqref{eq:Mobility}.  
A many sphere-grand mobility matrix is formed using the Stokesian dynamics method as in Ref \cite{durlofsky_brady_bossis_1987}.  
From there, this many sphere mobility matrix is condensed to a many-body grand mobility matrix by applying the conditions given by Eqs. \eqref{eq:RigidVel} and \eqref{eq:QuasiStatic} to each of the bodies present in the simulation.  
This step implicitly takes into account the internal constraining forces and torques acting on the spheres, which are necessary to maintain the rigid body conditions.   
To avoid additional complexities, all torques, forces and velocities are considered in the spatially-fixed frame of reference.  
With the torques acting on each body computed from Eq. \eqref{eq:Torque} and the external forces assumed to be zero, this many-body mobility matrix may be used to compute the translational velocity and rotational dynamics of each of the bodies.  
These velocity expressions are then integrated numerically using \texttt{MATLAB}'s built-in variable order differential equation solver \texttt{ode113}.  

In this work, each swimmer body is approximated as a permanent magnetic dipole, driven by a torque generated through an externally applied magnetic field.  In general, when two magnetic dipoles are present, an interaction force will arise between them.  In the simulations shown here, these dipole-dipole interaction forces are neglected, as the swimmer velocities resulting from these forces are negligible relative to those resulting from the magnetic torque due to the external field and from hydrodynamic interaction at the swimmer separation distances considered herein.  
In general, the force resulting from the interaction between two dipoles is of magnitude
\begin{equation}
    F_m = \frac{\mu_0m_1m_2}{4\pi r_{12}^4}
\end{equation}
where $\mu_0$ is the permeability of free space, $m_1$, $m_2$ are the dipole strengths, and $r_{12}$ is the magnitude of the position vector separating the dipoles  \cite{Furlani_magnetism}. 
For a swimmer separation distance of 4 spherical diameters, the velocity $V_m$ resulting from the magnetic interaction force can be shown to have magnitude $V_m/V^* = 8.75\times10^{-4}$ while the velocity $V_h$ due to hydrodynamic interaction between the swimmers has magnitude $V_h/V^* = 2.88$.
This separation distance is representative of the separations considered throughout this paper, so due to this difference in magnitude we neglect the forces arising from the dipole-dipole interaction.  

Also, in order for the mobility matrix derived from the Stokesian dynamics method to be considered exact at very small separation distances, it is necessary to either include an infinite number of force multipoles in the expansion of Eq. \ref{eq:multipole} or to include the near-field lubrication interactions by pairwise addition to the grand resistance matrix as explained by Durlofsky \emph{et al.} \cite{durlofsky_brady_bossis_1987}.  
However, these lubrication forces only become significant when there is relative motion between two spheres separated by a very small distance (e.g. less than 0.05 D*) \cite{durlofsky_brady_bossis_1987,brady_afm_1988,SwanBrady_Teaching,ishikawa_locsei_pedley_2008}.
In the simulations presented here, the spheres of each body are fixed, so there is no relative motion between them.  Further, all bodies in the simulations maintain separations large enough that these near-field forces do not become significant.  
For these reasons, lubrication forces are not considered in the model presented here. 

\begin{figure}
    \centering
    \begin{minipage}{.49\linewidth}
        \includegraphics[width=\linewidth]{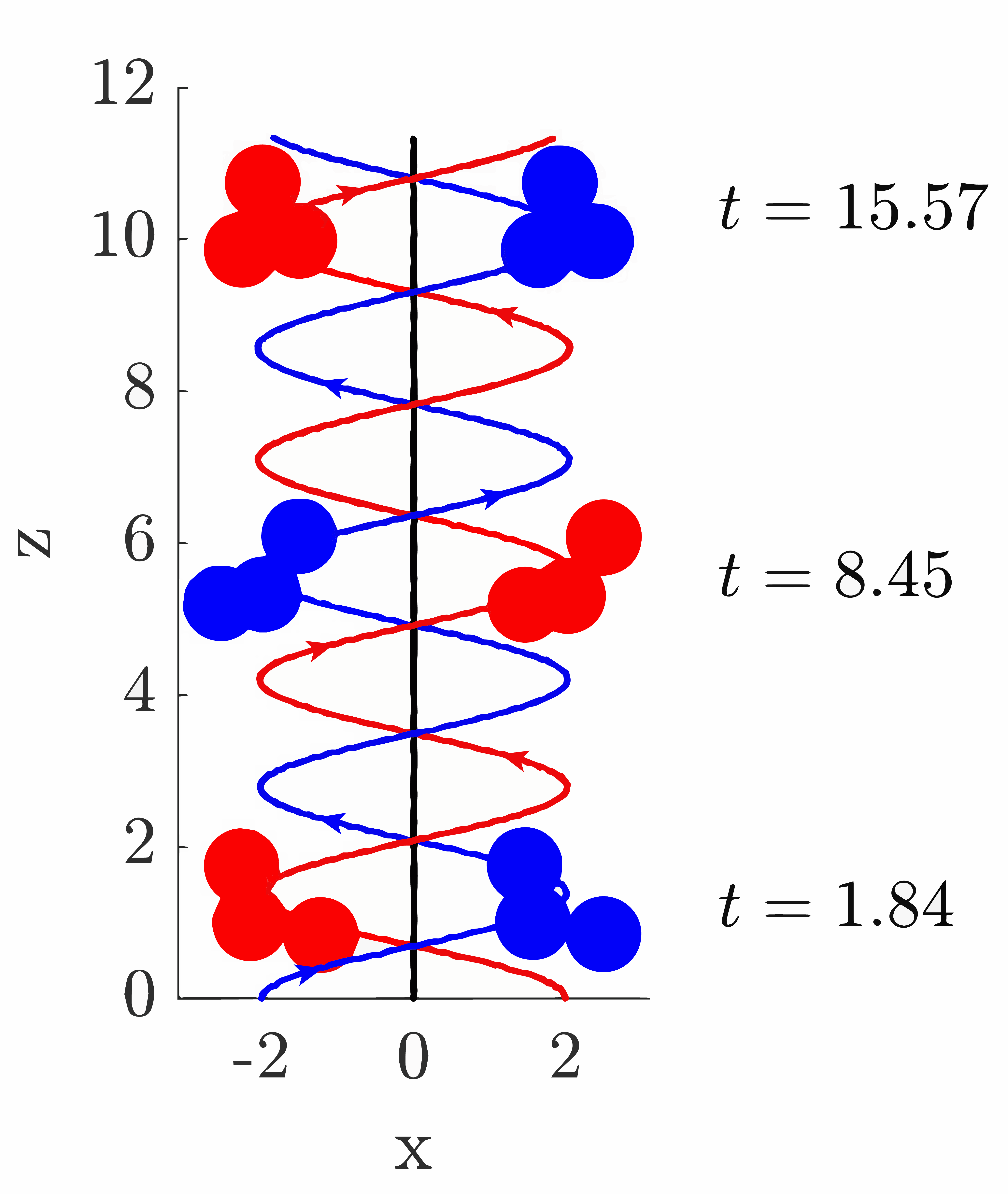}
    \end{minipage}
    \begin{minipage}{.49\linewidth}
        \centering
        \includegraphics[width=\linewidth]{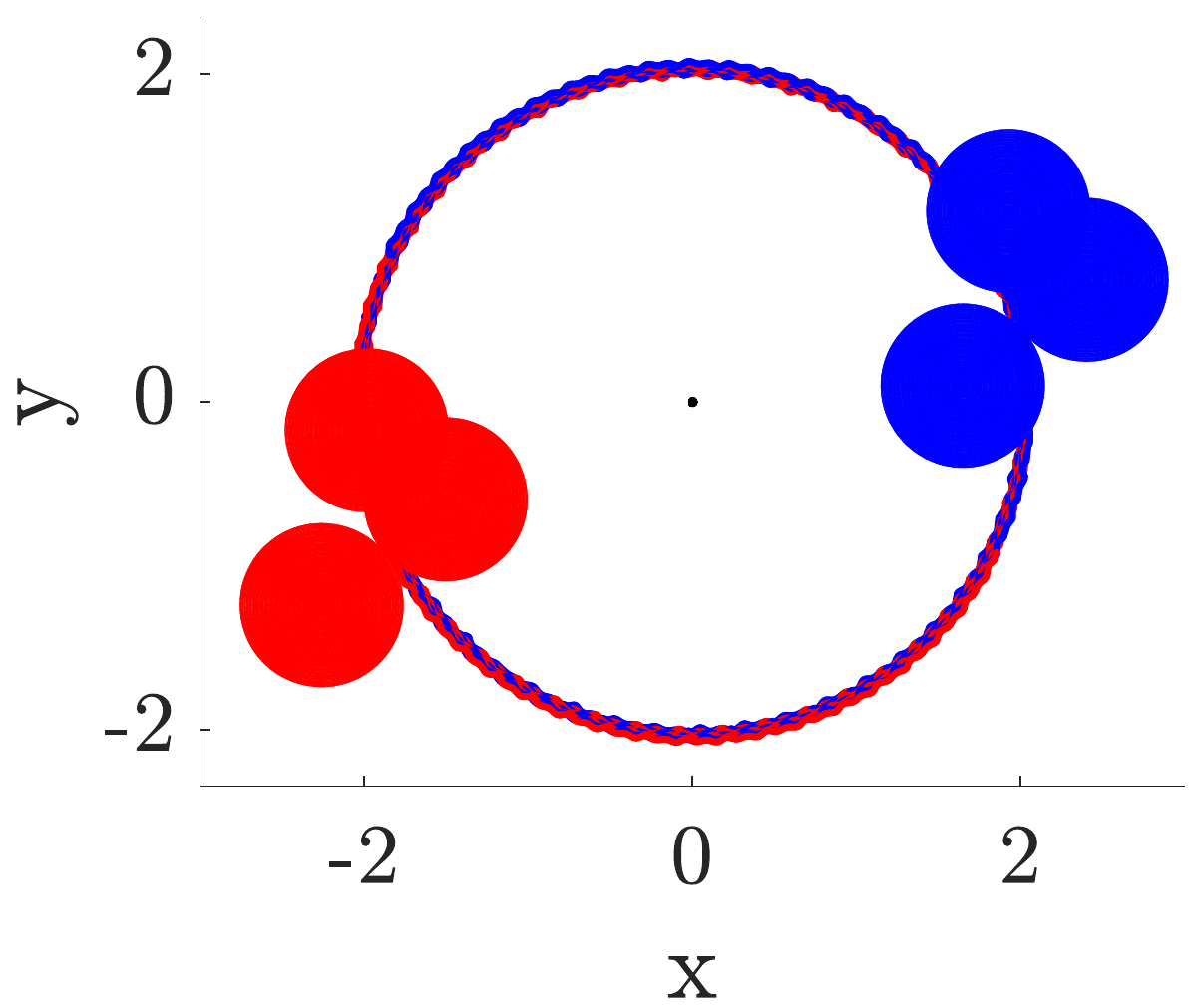}
    \end{minipage}
        \caption{Trajectory of two interacting microswimmers resulting from a rotating magnetic field with an axis of rotation perpendicular to the swimmer plane.  Initial swimmer separation is 4 spherical diameters. The black line indicates the path of the mean position of the two swimmers. }
    \label{fig:TwoSwimOoP}
\end{figure}

\subsection{Rotation orthogonal to swimmer plane}
Here two distinct classes of the trajectories of the interacting swimmers will be studied: (1) where the axis of rotation of the magnetic field is orthogonal to the plane initially containing the swimmers and (2) where the axis of rotation of the magnetic field is aligned with the line joining the swimmers.  
Fig. \ref{fig:TwoSwimOoP} shows the resulting swimmer trajectories associated with the first case, where the magnetic field's axis of rotation is orthogonal to the plane containing the two swimmers.
It can be seen that in this case, the swimmers tend to spiral about one another while translating in the direction of the magnetic field's axis of rotation. 
A similar trajectory was reported in the experimental work of Cheang \emph{et al.} \cite{Kim_APL2014} when two microswimmers passed each other swimming in opposite directions and spiraled about one another. 
That work reported that the instant of this hydrodynamic interaction corresponds to the peak speed of the swimmers, indicating that the effects of the hydrodynamic interaction, as seen here, are quite significant. 
The swimmers shown in Fig. \ref{fig:TwoSwimOoP} are initially spaced at a distance of four spherical diameters and this separation distance exhibits only very small oscillations about that constant value throughout the simulation.  
 
 \begin{figure}
    \centering
    \includegraphics[width=0.9\linewidth]{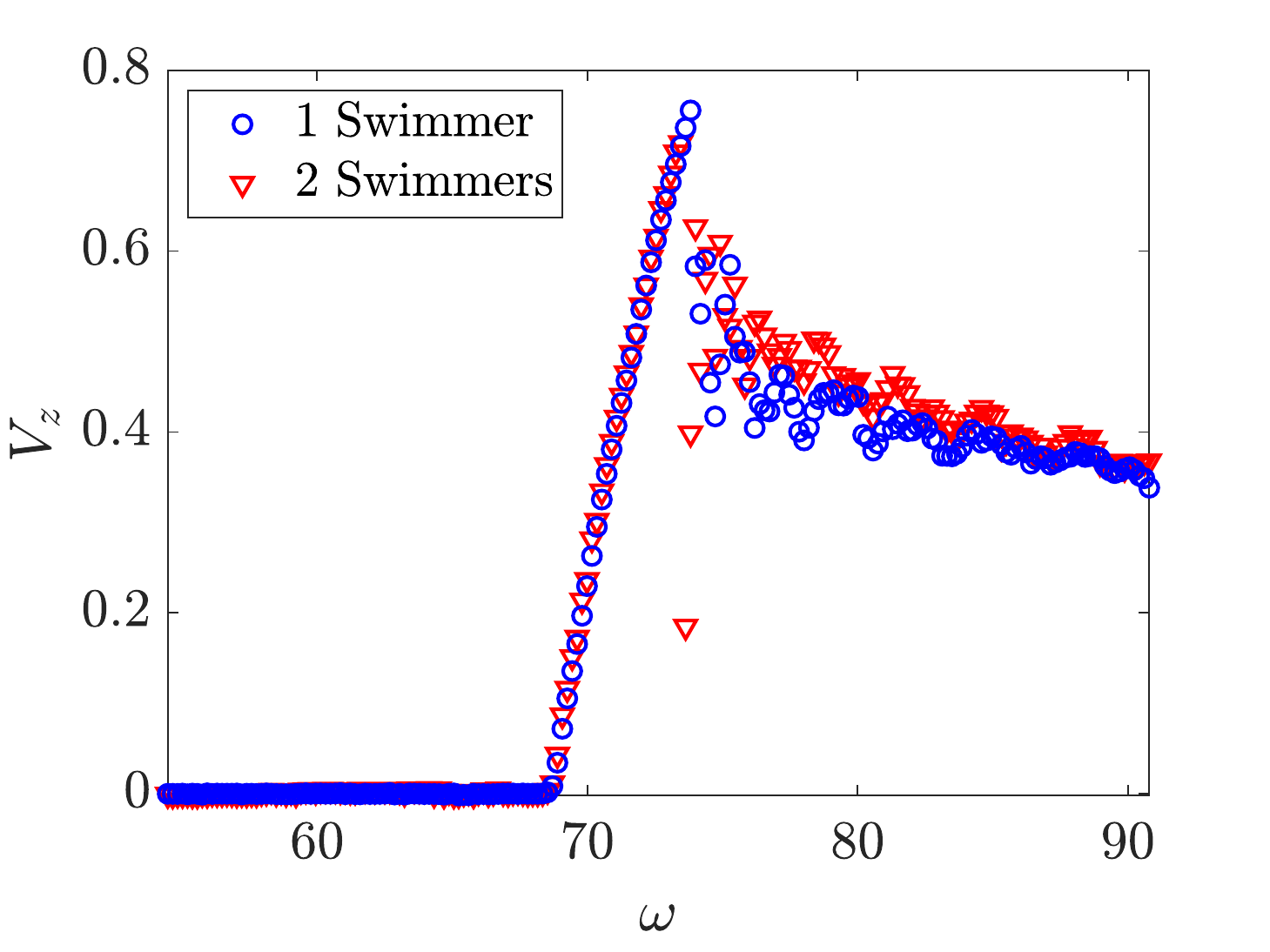}
    \caption{Comparison of the propulsion velocity of two interacting swimmers to a single swimmer.}
    \label{fig:VelocityCompare}
\end{figure}

Fig. \ref{fig:VelocityCompare} shows a comparison of the propulsion velocity of the two interacting swimmers to the propulsion of a single swimmer as the frequency of rotation $\omega$ of the magnetic field is varied. 
For the case of a single swimmer, the propulsion velocity remains near zero until the rotations reaches a critical frequency.  
Above this critical frequency, the velocity increases linearly with frequency until a second critical frequency, known as the \emph{step-out} frequency is reached.  
Beyond this step-out frequency, the propulsion velocity steadily begins to decay.  
This is a well known result that has been shown experimentally and theoretically for this geometry and many others (see for example \cite{kim_pre_2014,ManLauga,Leshansky_prf_2017}). 
It should be noted that this relationship only takes this specific shape for this particular choice of the magnetic moment vector.  
For other magnetic moment orientations, for example, one with a nonzero component in the body-frame $x$-direction, may show a differently shaped frequency-propulsion curve.  
This has been analyzed in detail by Morozov \emph{et al.} \cite{Leshansky_prf_2017}.  
For the case of two swimmers, the propulsion velocity is not affected greatly, aside from a few values lying very close to the step-out frequency.  
However, the magnitude of the velocity increases with the presence of a second swimmer due to the circular motion in the $xy$-plane that is not seen in the case of a single swimmer. 

In order to understand the spiraling pattern seen in the trajectory of the interacting swimmers, one only needs to study the fluid velocity field produced by the swimmers.  
In Fig. \ref{fig:VelField1}, it was shown that a single swimmer produces a velocity field that is highly rotational in planes orthogonal to the swimmer's axis of rotation.  
Furthermore, it was shown that of planes intersecting the swimmer, the magnitude of the fluid velocity field is much larger in the plane orthogonal to the magnetic field's axis of rotation than in those containing this axis.  
When considering this, it can be  inferred that this rotational fluid velocity field would tend to push a second  swimmer along a trajectory that is circular in this plane. 
Also, since the swimmers considered here are of identical geometry and magnetization, it should be expected that the trajectories of the two swimmers would be nearly identical.  

\begin{figure}[]
    \centering
    \includegraphics[width=0.9\linewidth]{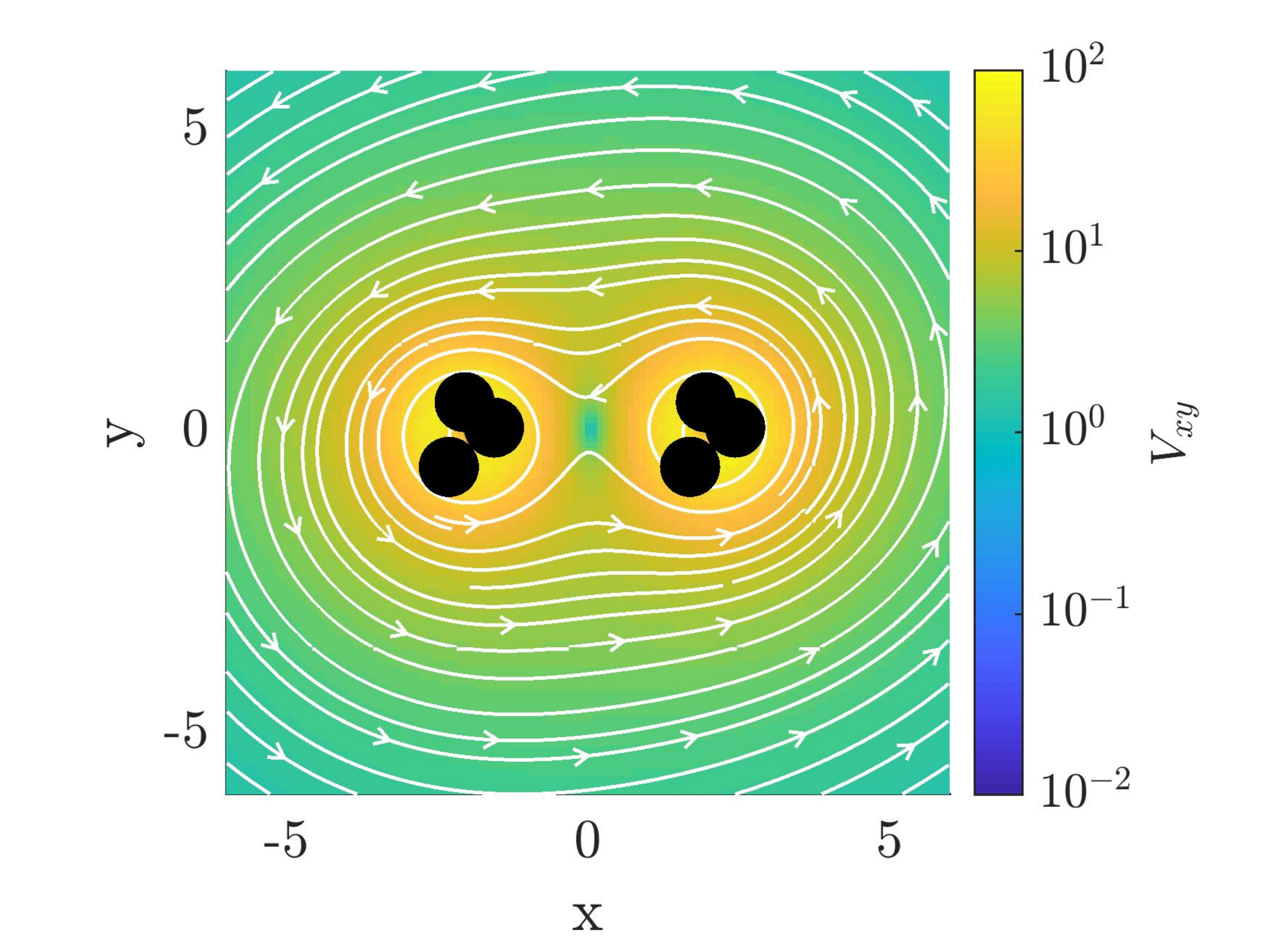}
    \caption{Streamlines and magnitude of fluid velocity field produced by two interacting swimmers, time averaged over a period of the magnetic field rotation. The velocity field is shown the plane orthogonal to the magnetic field axis of rotation. }
    \label{fig:VelField2}
\end{figure}

These trajectories are further explained by studying the fluid velocity field generated by the two swimmers in the plane orthogonal to the swimmer axes of rotation.  
The streamlines and magnitude of this velocity field are shown in Fig. \ref{fig:VelField2}.  
Again, this velocity field is time-averaged over the period of the magnetic field's rotation.  
This time-average is computed by first integrating the swimmer velocities using the Stokesian dynamics procedure to obtain the swimmer positions and orientations over time.  
With this, the fluid velocity field is computed as the swimmer bodies are rotated accordingly, but with their positions remaining fixed. 

It is insightful to compare the velocity field produced by the swimmers to that of a rotlet, a point torque singularity.  
The fluid velocity at a point $\mathbf{x}_E$ produced by a rotlet located at a point $\mathbf{x}_R$ may be written as \cite{pozrikidis_1992,kim_karrila}
\begin{equation}
    \mathbf{u}_R(\mathbf{x}_E) = \frac{1}{8\pi\mu}\frac{\mathbf{T}\times\mathbf{r}_{EP}}{r_{EP}^3}
\end{equation}
where $\mathbf{r}_{EP} = \mathbf{x}_E - \mathbf{x}_R$, $r_{EP} = \|\mathbf{r}_{EP}\|$, and $\mathbf{T}$ is the vector giving the magnitude and direction of the point torque. 

\begin{figure}[]
    \centering
    \includegraphics[width=0.8\linewidth]{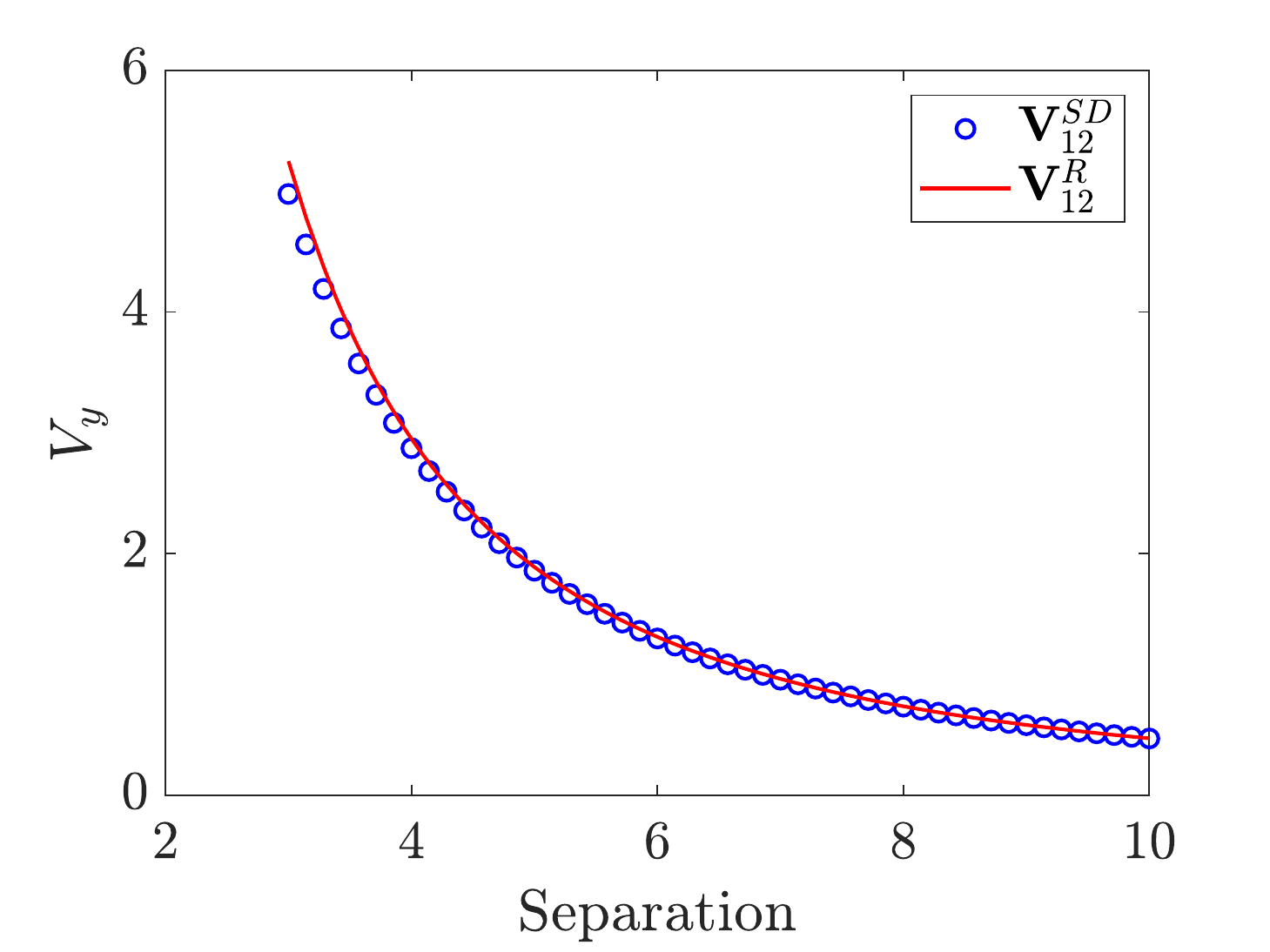}\\
    \includegraphics[width=0.8\linewidth]{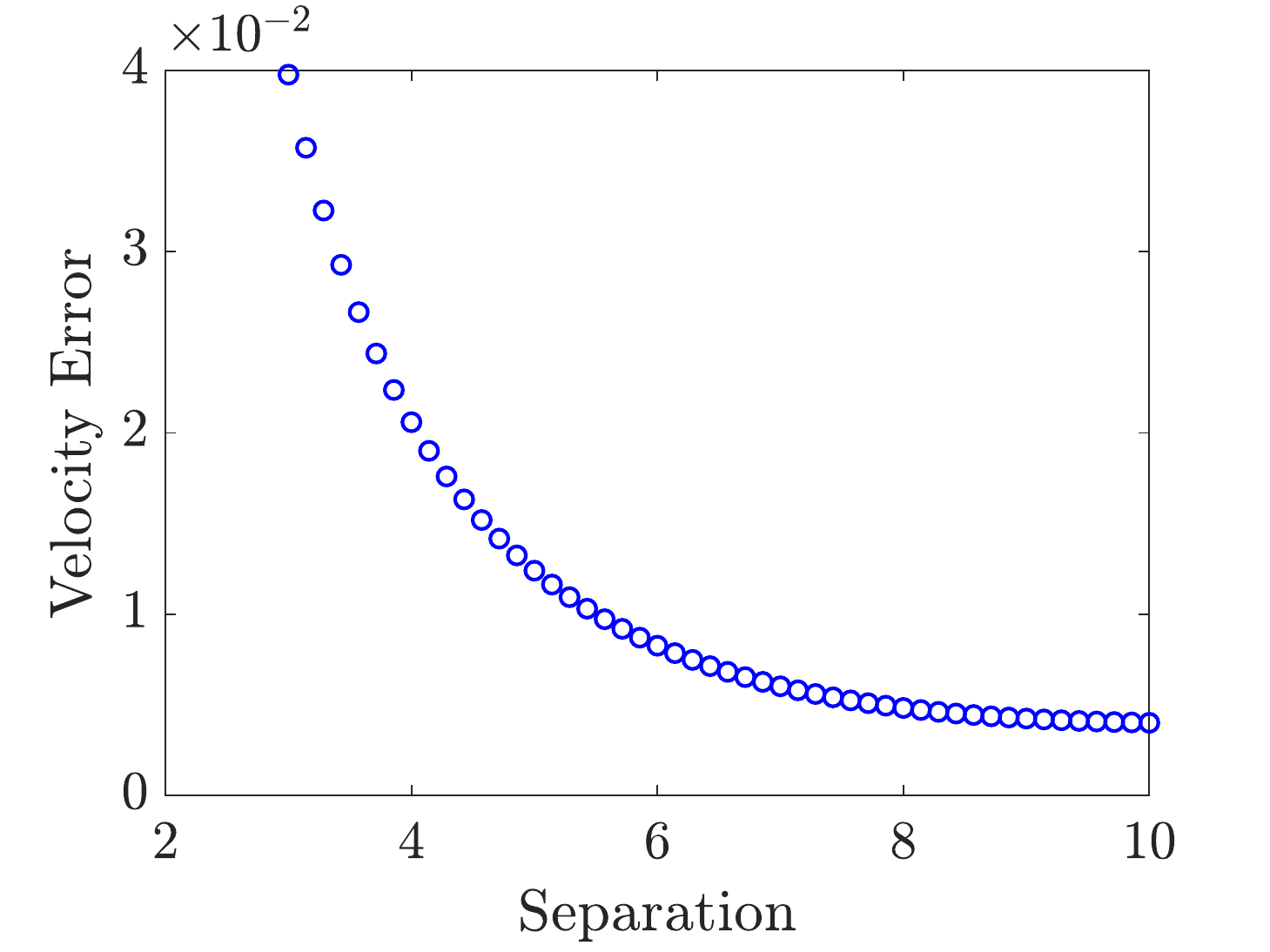}
    \caption{Comparison of velocity of swimmer 1 due to hydrodynamic interactions with swimmer 2 as predicted using the many-body grand mobility matrix of the Stokesian dynamics procedure and the rotlet velocity field approximation.}
    \label{fig:RotletApprox}
\end{figure}

From the Stokesian dynamics approach for computing the interactions as described above, the velocity of one of the two swimmers is written as 
\begin{equation}
    \begin{pmatrix}
    \mathbf{V}_1\\
    \boldsymbol{\Omega}_1
    \end{pmatrix}
    =
    \MM_{11}
    \begin{pmatrix}
    \mathbf{F}^e_1\\
    \mathbf{T}^e_1
    \end{pmatrix}
    +
    \MM_{12}
    \begin{pmatrix}
    \mathbf{F}^e_2\\
    \mathbf{T}^e_2
    \end{pmatrix}
\end{equation}
where the notation for the mobility matrix has been introduced such that
\[
\MM_{ij} =     
    \begin{pmatrix}
    \mathbf{K}_{ij}  &   \mathbf{C}_{ij}\\
    \mathbf{C}^{\intercal}_{ij}    &   \mathbf{M}_{ij}
    \end{pmatrix}.
\]
In this expression, the first term represents the hydrodynamic self-mobility, as given in Eq. \eqref{eq:Mobility}, and the second term takes into account the effect of hydrodynamic interaction with the second swimmer. 
To approximate the swimmer-swimmer interactions using the rotlet velocity field, one may consider swimmer 2 as a rotlet and evaluate the fluid velocity field at the location of swimmer 1. 
That is,
\begin{equation}
    \mathbf{V}_{12}^R = \frac{1}{8\pi\mu}\frac{\mathbf{T}_2^e\times\mathbf{r}_{12}}{r_{12}^3}.
\end{equation}
To check the validity of this approximation, it is compared to the interaction velocity produced by the interaction mobility matrix $\MM_{12}$, as predicted by the Stokesian dynamics algorithm, 
\begin{equation}
    \mathbf{V}_{12}^{SD} = \mathbf{C}_{12}\mathbf{T}_2^e
\end{equation}
The results of this calculation are shown in Fig. \ref{fig:RotletApprox}  (a) for the swimmer configuration as shown in Fig. \ref{fig:VelField2} above, where the rightmost swimmer is denoted as swimmer 1. 
In this configuration, both swimmers lie on the $x$-axis, so the tendency of swimmer 1 is to move in the positive $y$ direction at that instant.  
Fig. \ref{fig:RotletApprox} (b) shows a normalized error in the rotlet calculation, given by $\|\mathbf{V}_{12}^R - \mathbf{V}_{12}^{SD}\|/\|\mathbf{V}_{12}^{SD}\|$.  
These results indicate that the rotlet approximation given above is a good one, even at quite close separation distances.  
This is somewhat surprising when considering that the rotlet approximation effectively neglects the finite size of both swimmers, assuming that swimmer 2 can be represented by a single point torque and that swimmer 1 behaves just as a fluid particle located at its position would.  

This finding is quite useful, as it gives a very low order approximation for the hydrodynamic interaction of two magnetic swimmers that allows one to qualitatively predict the dynamics of multiple interacting swimmers. 
The dynamics of interacting rotlets are  better understood \cite{marchetti_sm_2012,LeoniLiverpool2010EPL, lushi_jnls_2015} and easier to incorporate into control models \cite{bft_ijira_2018}.
This could also allow for significant computational advantage with increasing number, $N$, of swimmers  since the simulation of $N$ interacting rotlets is significantly faster than the Stokesian dynamics approach.
Further, the qualitative understanding provided by the rotlet approximation yields a simple explanation of the helical trajectory of the interacting swimmers. 
It is well known \cite{marchetti_sm_2012,LeoniLiverpool2010EPL,lushi_jnls_2015} that two rotlets with identical torques which are perpendicular to the plane containing them tend to move in a circular trajectory in that plane. 
For the swimmers considered here, the torques are not necessarily perpendicular to the plane containing them at any instant, but due to the rotational nature of the swimmers, the components of the torque in the plane containing the swimmers average to nearly zero over the time period of the rotations. 
This explains the cooperative, circular motion in the plane containing the microswimmers.  
As seen in Fig. \ref{fig:VelField1}, the fluid velocity field decays very rapidly in the planes containing the swimmer's axis of rotation.  
Therefore, the motion of the swimmers in the direction of rotation of the magnetic field is negligibly different from the case of just a single swimmer, as further evidenced by Fig. \ref{fig:VelocityCompare}.  
Thus, the helical trajectory of the interacting swimmers may be thought of as the superposition of the trajectory of a single swimmer in free space with the trajectory of a rotor interacting cooperatively with another rotor in the plane orthogonal to the magnetic field's axis of rotation. 

\subsection{Rotation along line joining swimmers}

\begin{figure}[]
    \centering
    \begin{minipage}{.49\linewidth}
        \centering
        \includegraphics[width=\linewidth]{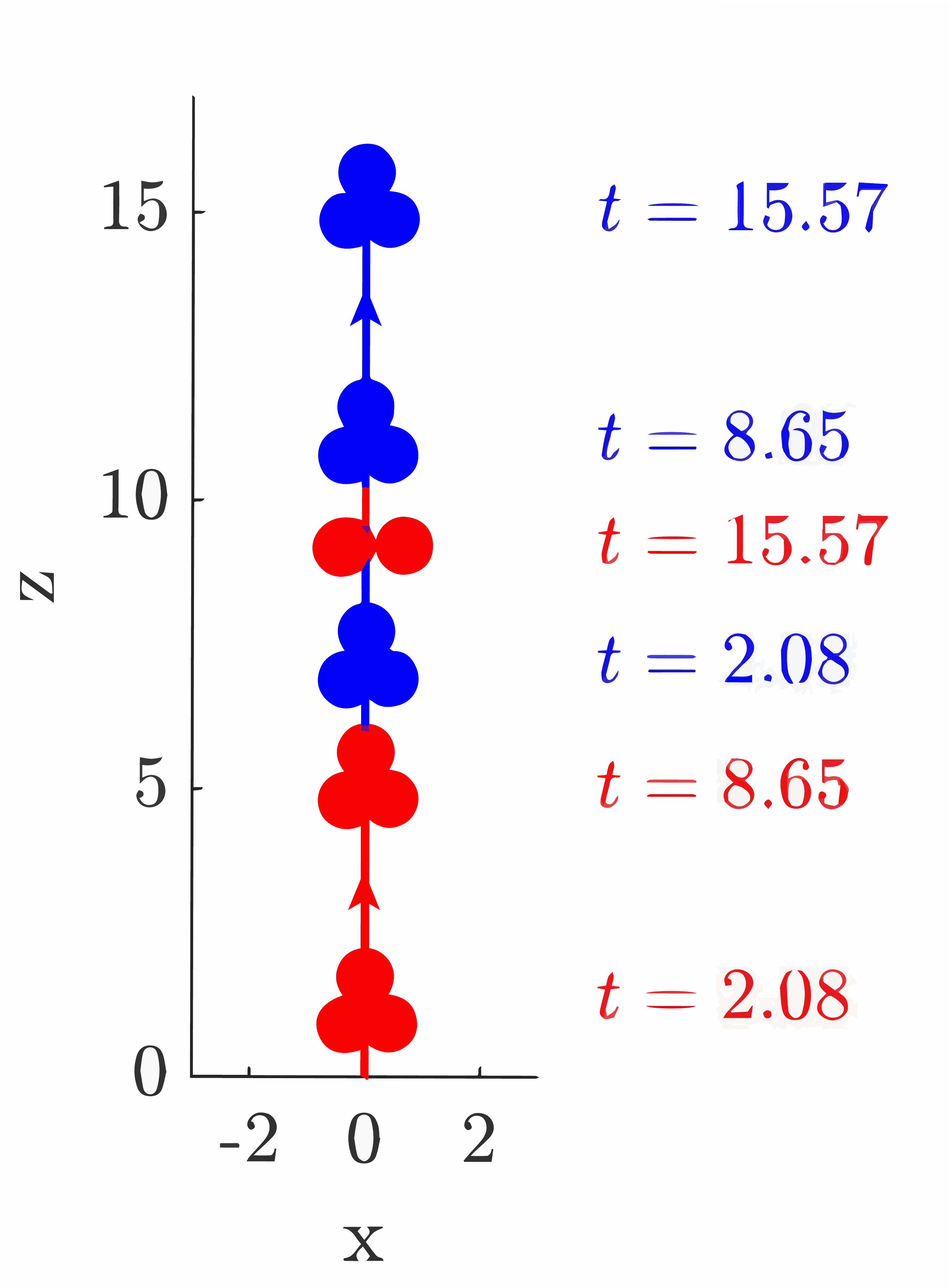}
    \end{minipage}%
    \begin{minipage}{.49\linewidth}
        \centering
        \includegraphics[width=\linewidth]{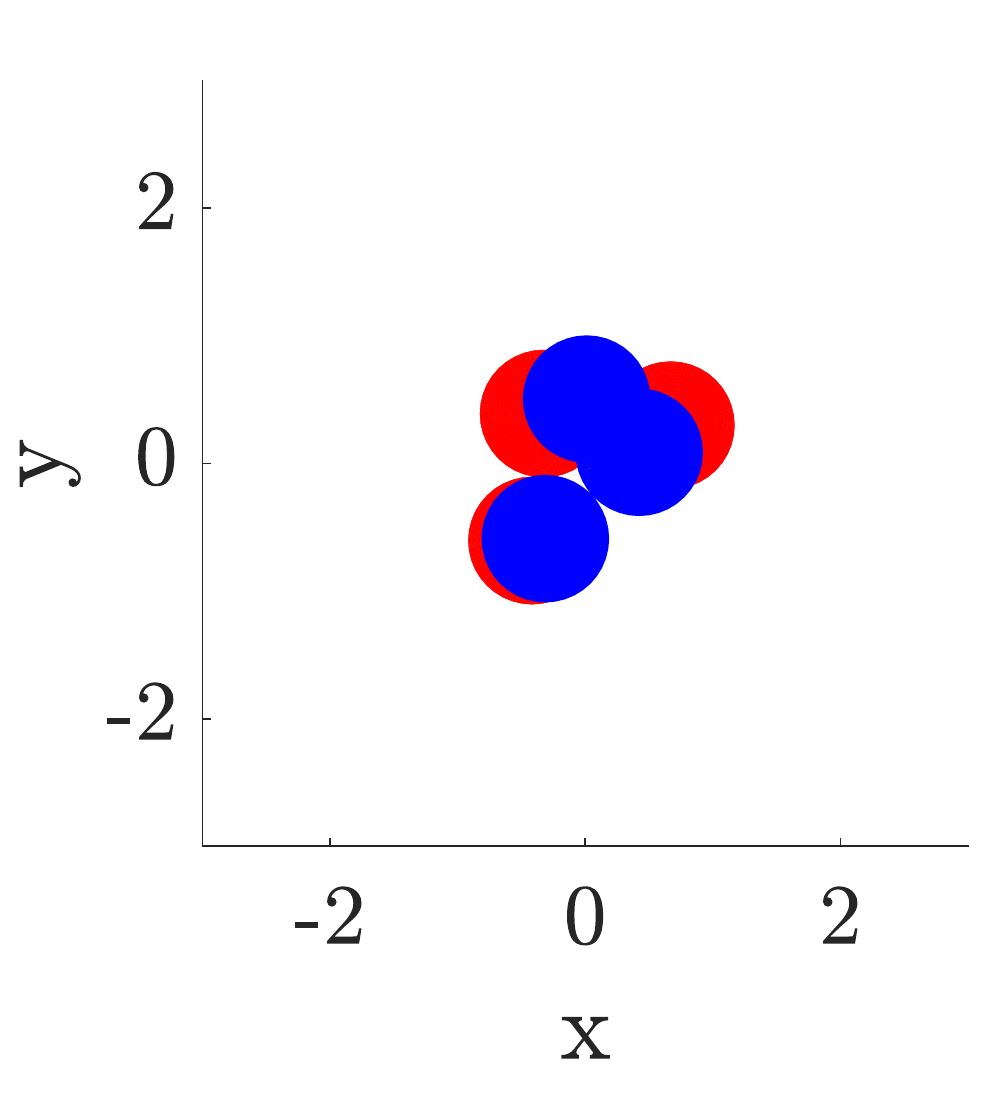}
    \end{minipage}%
        \caption{Trajectory of two microswimmers resulting from a rotating magnetic field with an axis of rotation aligned with the line initially separating the swimmers. Initial swimmer separation is 6 spherical diameters. }
    \label{fig:TwoSwimIP}
\end{figure}

A second significant case that must be considered in studying the interactions of two of these artificial microswimmers is the case where the axis of rotation of the magnetic field is aligned with the line joining the two swimmers. 
The resulting trajectories for this case, as computed by integrating the equations of motion using the Stokesian dynamics algorithm, are shown in Fig. \ref{fig:TwoSwimIP}. 
In this case, both swimmers move in the positive $z$ direction, along the direction of rotation of the magnetic field, with very little motion in the plane orthogonal to the magnetic field's axis of rotation. 

As before, this motion can be explained by considering the fluid velocity field produced by each of the swimmers and approximating it by the velocity field produced by a rotlet. 
For each of the bodies, the fluid velocity field produced will again be largely in the plane containing the swimmer and orthogonal to the axis of rotation.  
However, since the swimmers are initially aligned along their direction of rotation, this plane will not contain the second swimmer.  
Instead, the other swimmer lies in a plane containing the axis of rotation, in which the fluid velocity field was shown to decay very rapidly to negligibly small values moving radially away from the swimmer body in Fig. \ref{fig:VelField1}. 

Considering this configuration in terms of the rotlet model, the time averaged torque is still nearly aligned with the direction of rotation of the swimmer as before.  
Thus, the only nonzero component of the time-averaged torque will be in the $z$-direction.
Further, the only nonzero component of the $\mathbf{r}_{12}$ vector will also be in the $z$-direction.  
For these reasons, at any point lying on the axis of rotation, the velocity field produced by the rotlet should vanish.  
Therefore, when averaged over time, one can expect the influence of each swimmer on the other to be negligible, as is indicated by the trajectories in Fig. \ref{fig:TwoSwimIP}, in which the paths of the swimmers seem to differ very little from the case of a single swimmer in free space. 

This observation, along with the rotlet approximation also allow for a good qualitative prediction of what the motion of more than two swimmers will look like.  For these cases, one can predict that swimmer bodies that lie in the same plane orthogonal to the direction of rotation of the magnetic field will tend to interact, while swimmers with significant separation in the direction of rotation will have little influence on one another.  The groups of swimmers that experience significant hydrodynamic interaction tend to move in trajectories about their geometric center, though these trajectories may be irregular or non-symmetric depending on initial configuration.  In addition to this interaction, the swimmer bodies also experience a net motion in the direction of rotation for a certain range of driving frequency as shown in Fig. \ref{fig:VelocityCompare}.  Since neither of these motions result in separation of swimmers in the group, the swimmers tend to remain together in a coherent group, as long as their magnetizations lie within a range that leads to propulsion in the same direction.  This is demonstrated further for larger groups of swimmers in the simulations of the following section. 


\section{Larger groups of swimmers}

\begin{figure}[]
    \centering
    \includegraphics[width=0.49\linewidth]{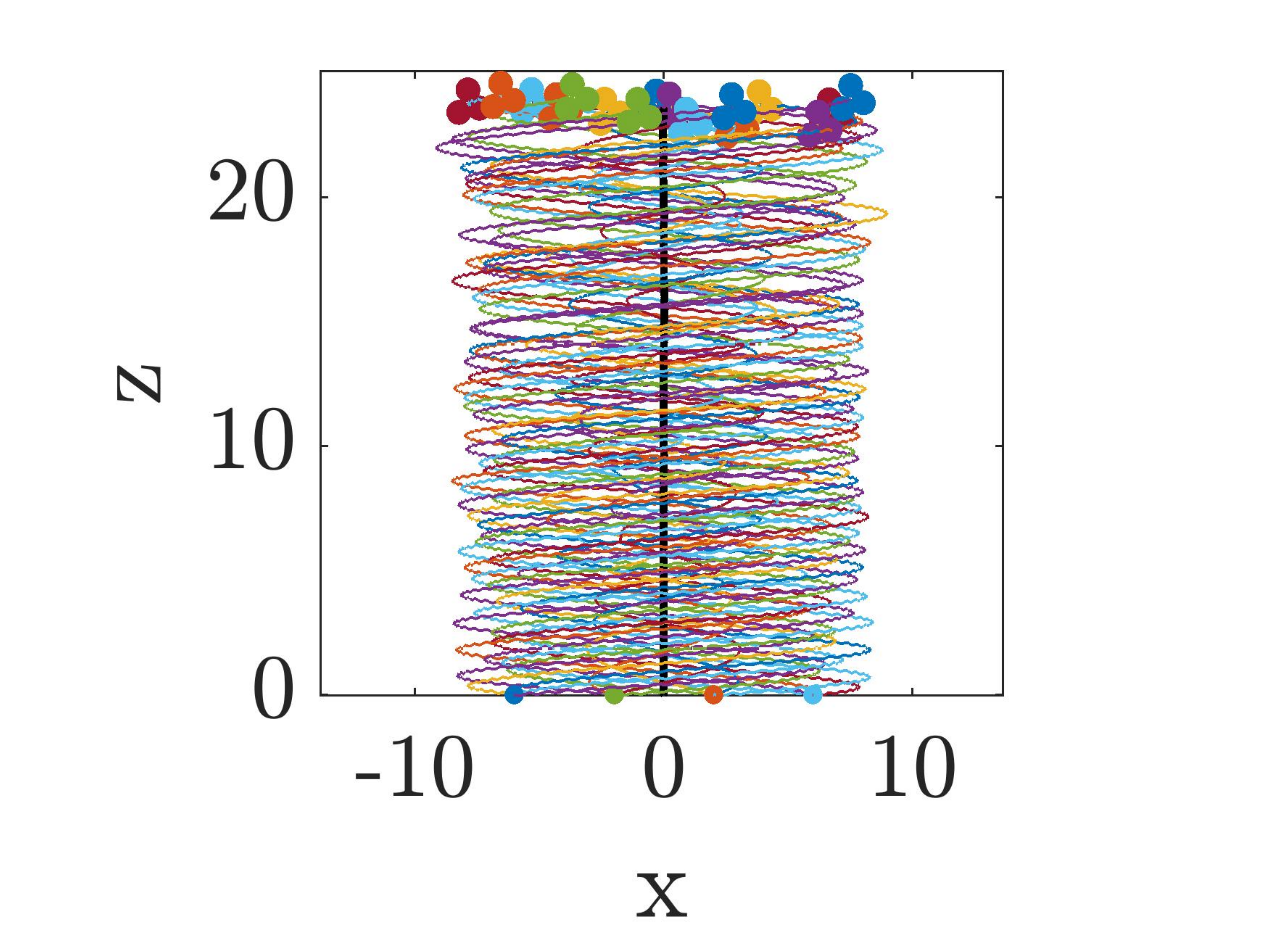}
    \includegraphics[width=0.49\linewidth]{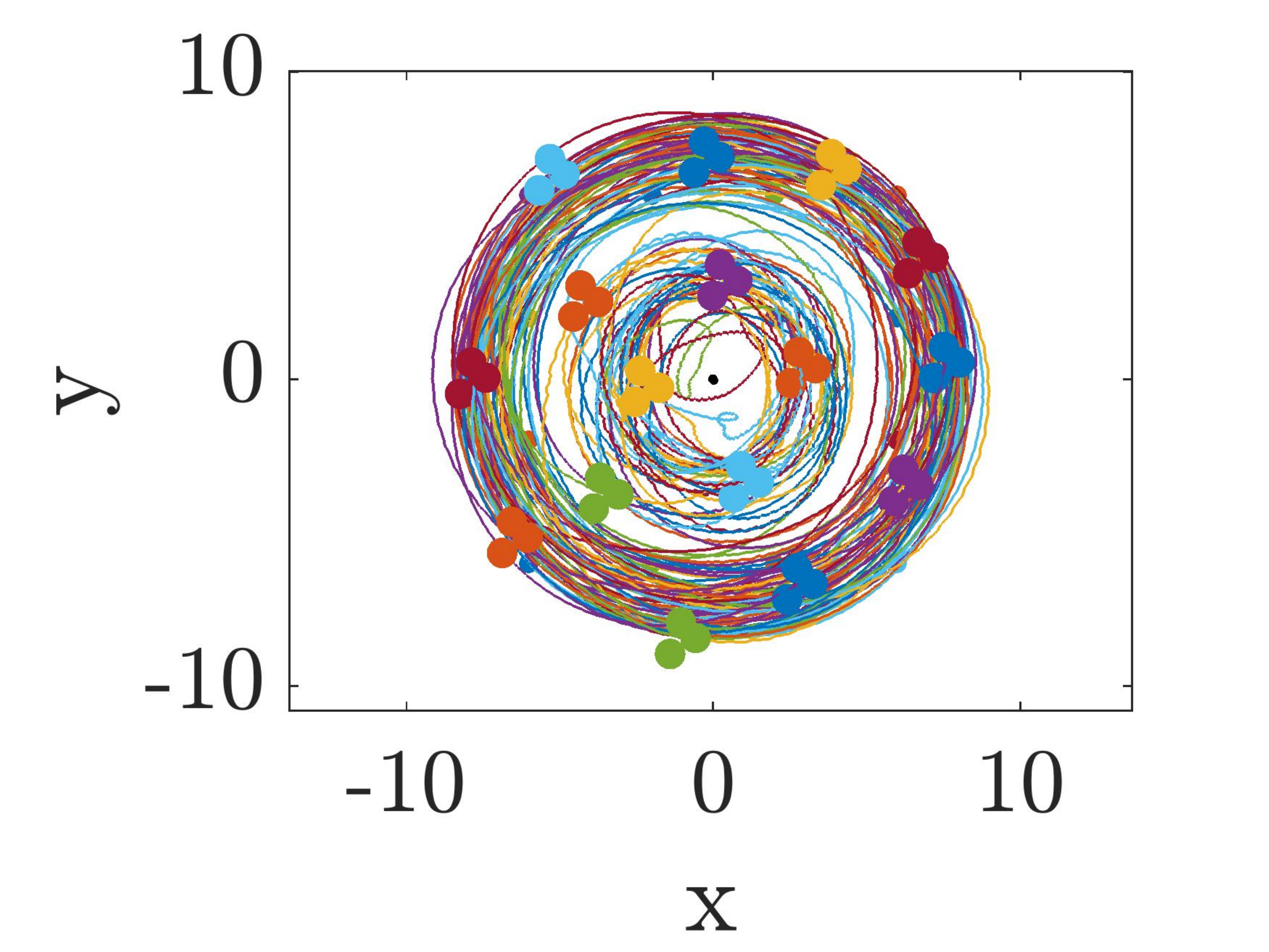}\\
    (a) \hspace{1.4in} (b)\\
    \includegraphics[width=0.49\linewidth]{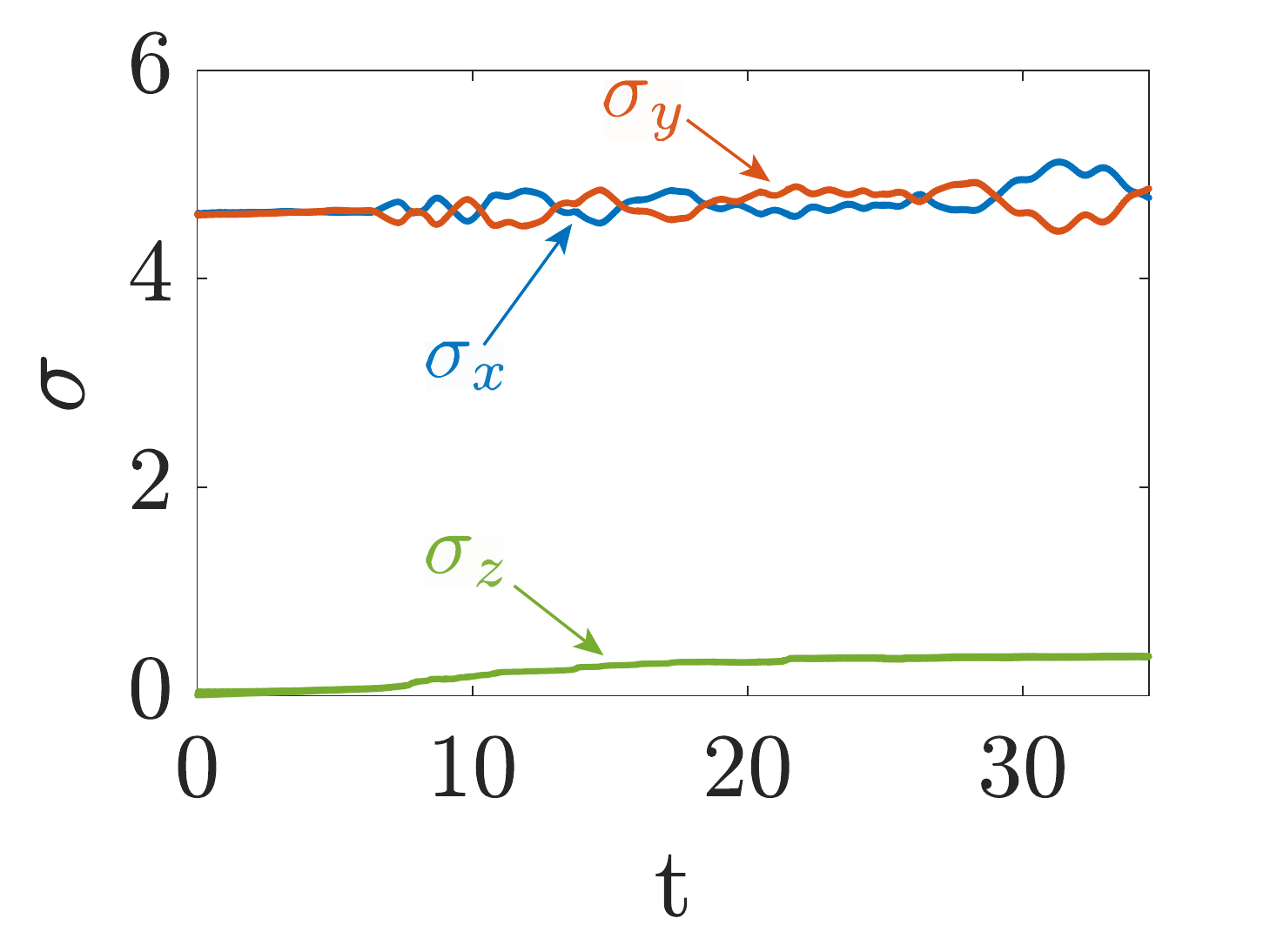}
    \includegraphics[width=0.49\linewidth]{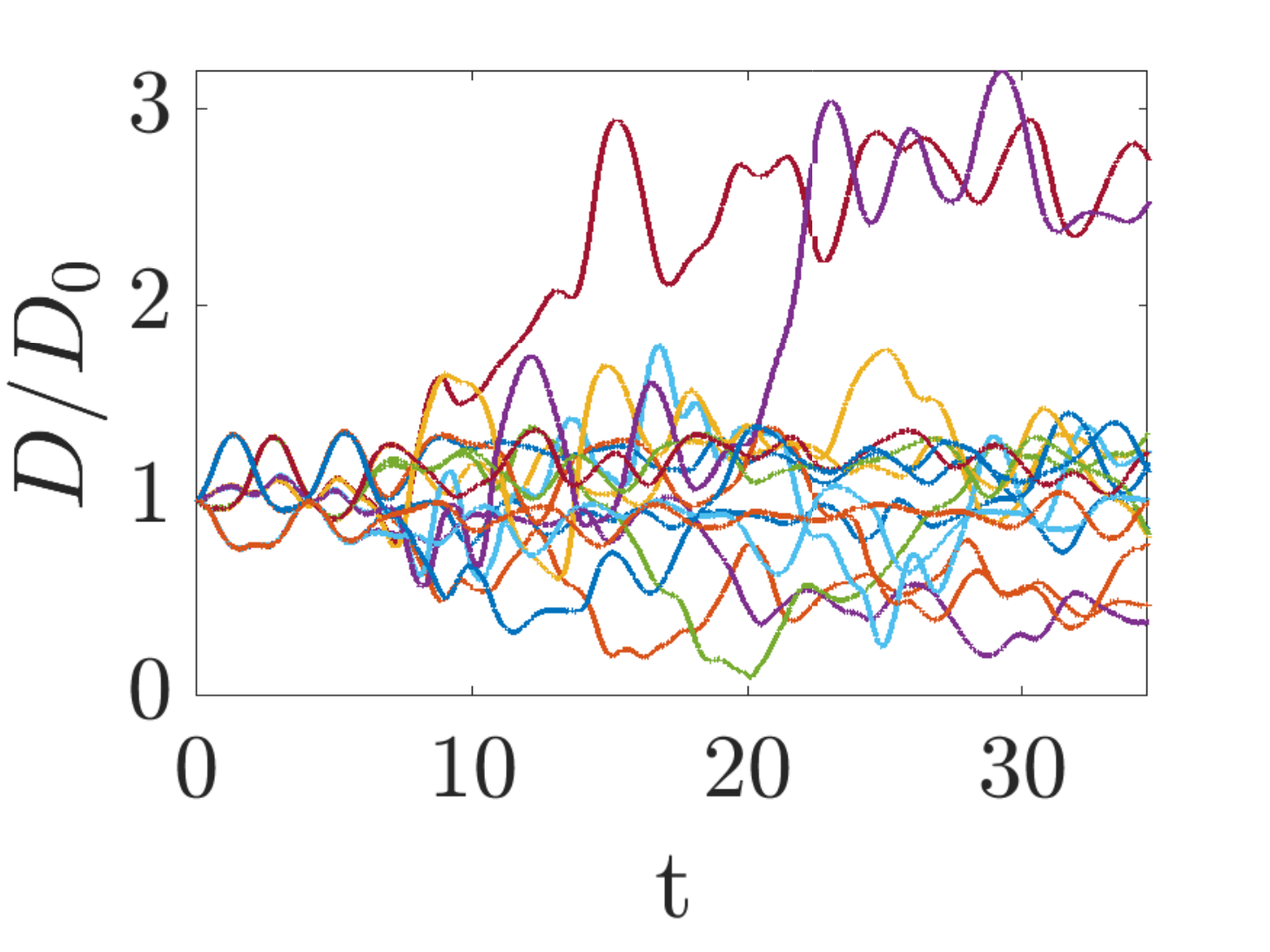}\\
    (c) \hspace{1.4in} (d)\\
    \caption{Simulation results for 16 identical interacting swimmers initially placed in a plane orthogonal to the magnetic field rotation direction.  (a),(b) give trajectories of the bodies in the $xy$- and $xz$- planes.  The black line indicates the path of the mean position of the swimmers. Filled circles indicate the starting position, while the swimmer bodies indicate the final position and orientation. (c) Standard deviation of the position coordinates from the mean position coordinate. (d) Distance of each swimmer from the group center of mobility as a ratio of the initial distance from this center.}
    \label{fig:16Swim}
\end{figure}

When considering larger groups of these swimmers in simulation, much of the dynamics can be explained by extending the reasoning of the two cases of two swimmers detailed above.  
As a first case, a uniform grid of sixteen swimmers is studied, with swimmers initially positioned in the $xy$-plane at a separation distance of 4 spherical diameters.  To start, a group of identical microswimmers is considered; that is, all swimmers have the same geometry and magnetic moment orientation. 
As before, a uniform magnetic field will be considered to be steadily rotating in the $xy$ plane, with its axis of rotation aligned with the spatial frame $z$-direction, as given in Eq. \eqref{eq:MagField}. 
The resulting trajectories of these bodies are shown in Fig. \ref{fig:16Swim} (a)-(b).  

In the case of two interacting microswimmers with a magnetic field rotating orthogonally to the plane containing the swimmers, it was shown that the swimmer trajectories were not significantly altered in the direction of the magnetic field rotation due to the interaction.  
However, the trajectories in the plane containing the swimmers varied, as the swimmers begin to move in circular paths in this plane due to hydrodynamic interactions.  
Similarly, in the case of 16 identical swimmers initially starting in a plane orthogonal to the magnetic field rotation, the swimmers tend move along a spiraling trajectory in the plane containing them, with only small displacement from this plane throughout the motion.  
The trajectory plots in Fig. \ref{fig:16Swim} (a)-(b) also depict the path of the mean position of the swimmers' centers of hydrodynamic mobility.
This mean position will be referred to as the group center of mobility in the remainder of this paper. 
It can be seen that this path is remarkably straight and aligned with the direction of rotation of the magnetic field, much like what was seen in the case of two swimmers and for a single swimmer.

The distortion of the volume occupied by the group of swimmers can be quantified by the standard deviations of the positions of  the swimmers from the mean position of all swimmer bodies in each of the $x$-, $y$-, and $z$- coordinate directions. 
These standard deviations thus give a measure of how much the size of the group of swimmers is expanding and contracting in the different spatial coordinate directions. 
The plot of these standard deviations over time is shown in Fig. \ref{fig:16Swim} (c).
From this, it is seen that there is only a small increase in the deviations in any of the $x$-, $y$-, or $z$- coordinates throughout the simulation.  
This shows that the swimmers tend to move as a coherent group around the mean trajectory, without significant distortion of the volume occupied by the swimmers. 
In Fig. \ref{fig:16Swim} (d), the  distances of each swimmer $D$, from the group center of mobility are shown as a ratio of the their initial distance, $D_0$, from the group center of mobility.  
From this, it is clear that most swimmers largely rotate in groups about the mean position of the group such that $D/D_0$ oscillates about one. 
However, a few swimmers stray from the mean, while a few move closer to the mean. 
These transitions from an inner region to the outer region and vice versa occur in such a way that the standard deviation of the positions of the swimmers does not change significantly.

\begin{figure}[]
    \centering
    \includegraphics[width=0.55\linewidth]{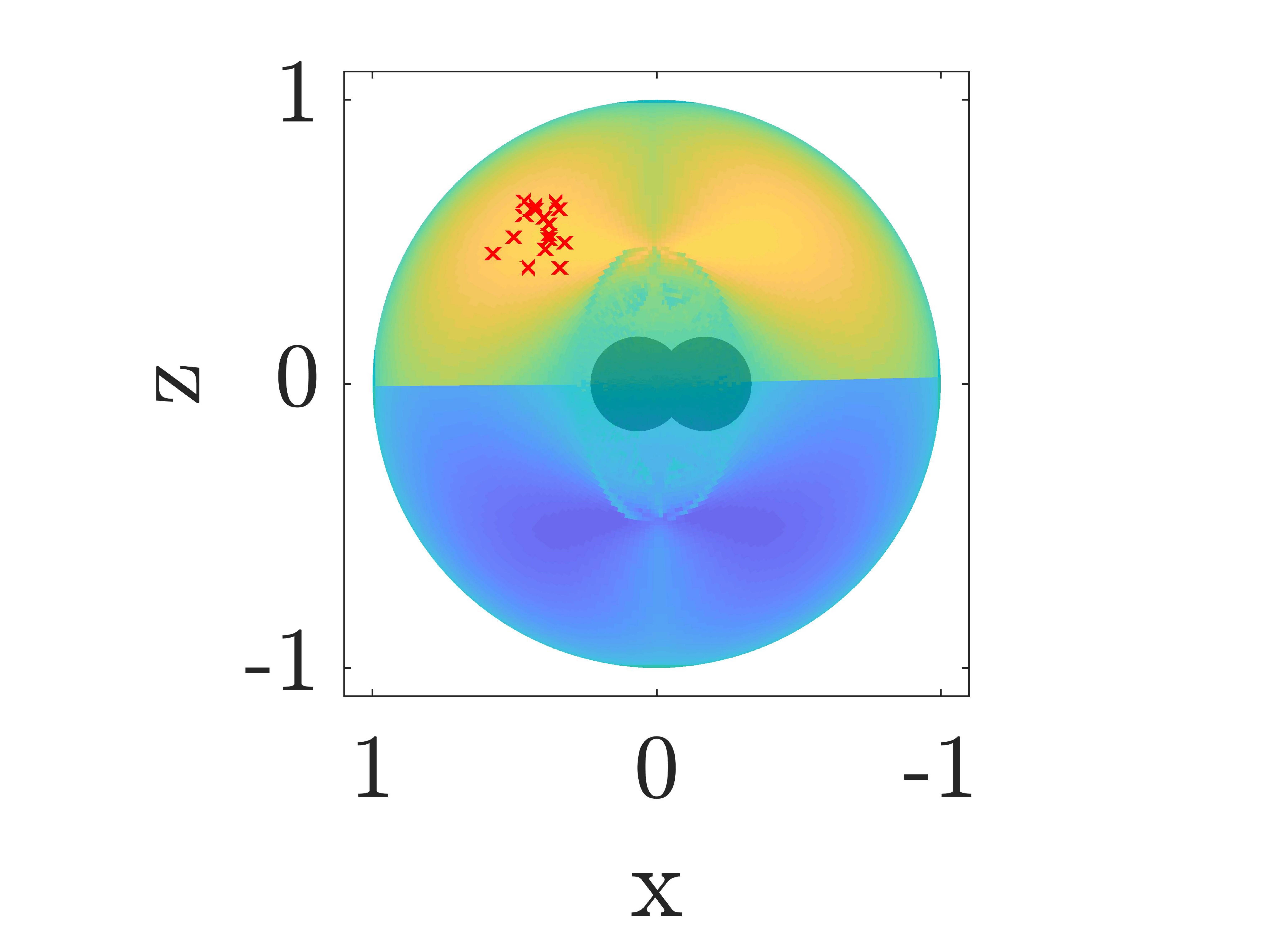}\\
    (a)\\
    \includegraphics[width=0.49\linewidth]{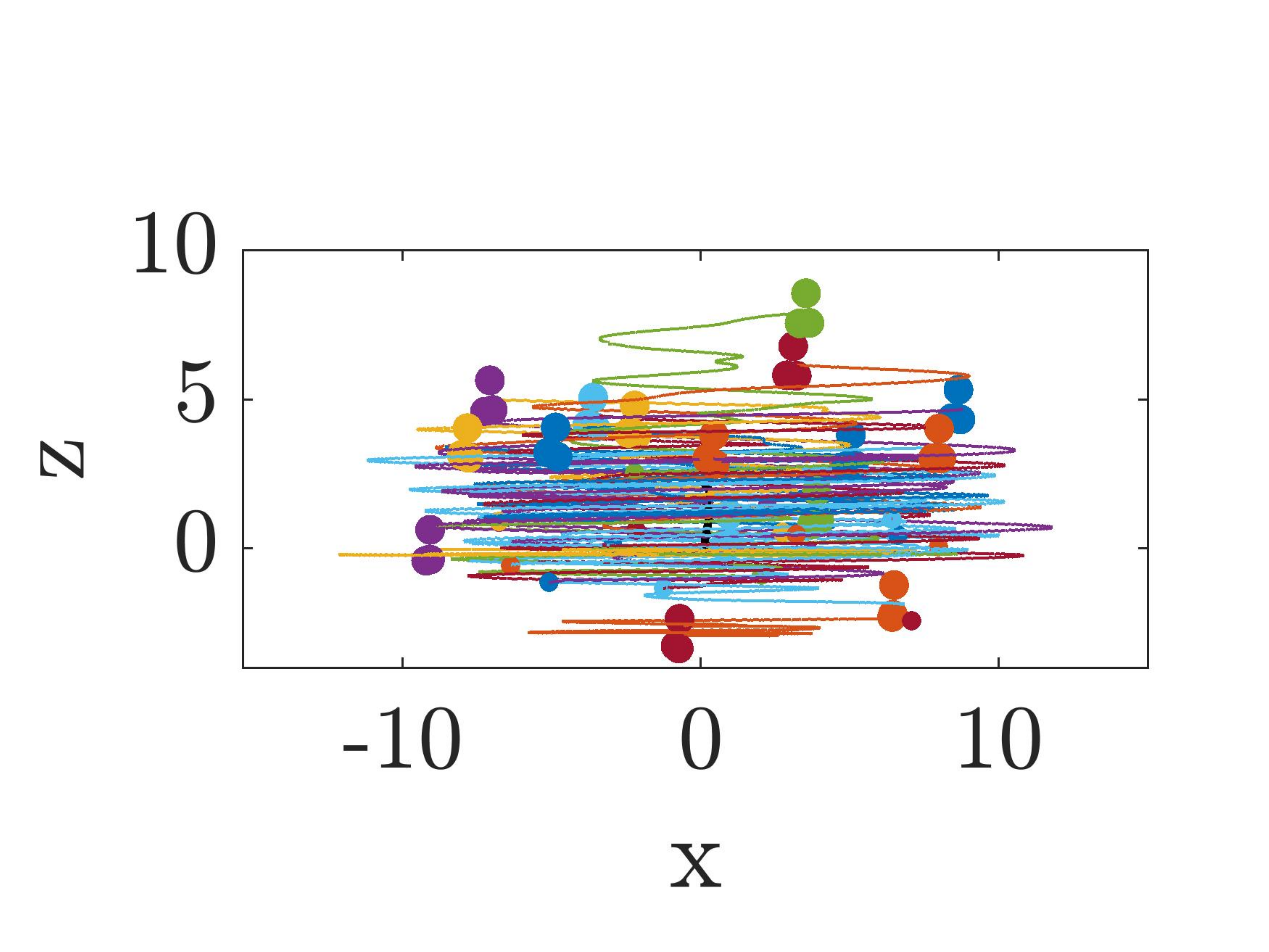}
    \includegraphics[width=0.49\linewidth]{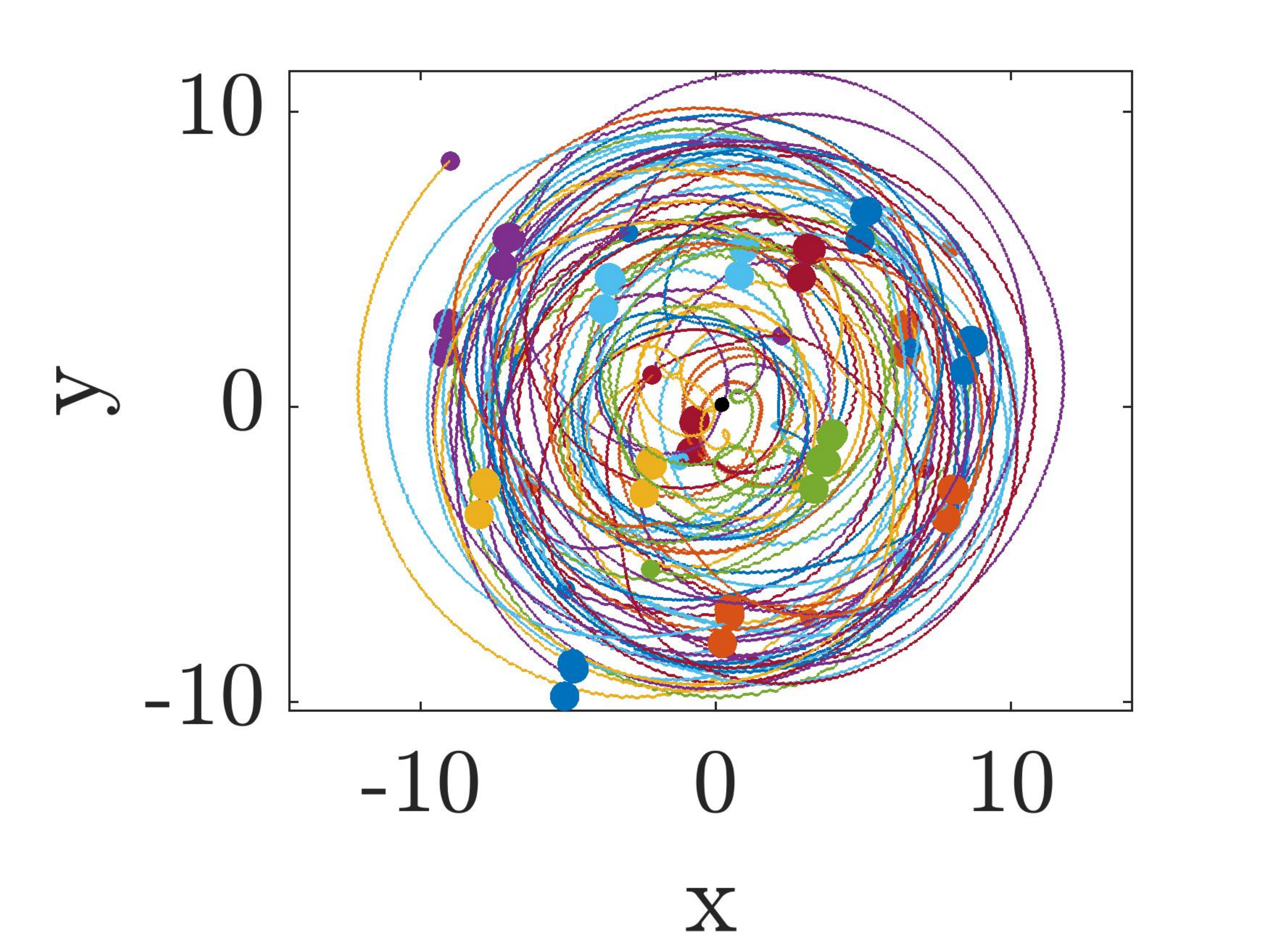}\\
    (b) \hspace{1.4in} (c)\\
    \includegraphics[width=0.49\linewidth]{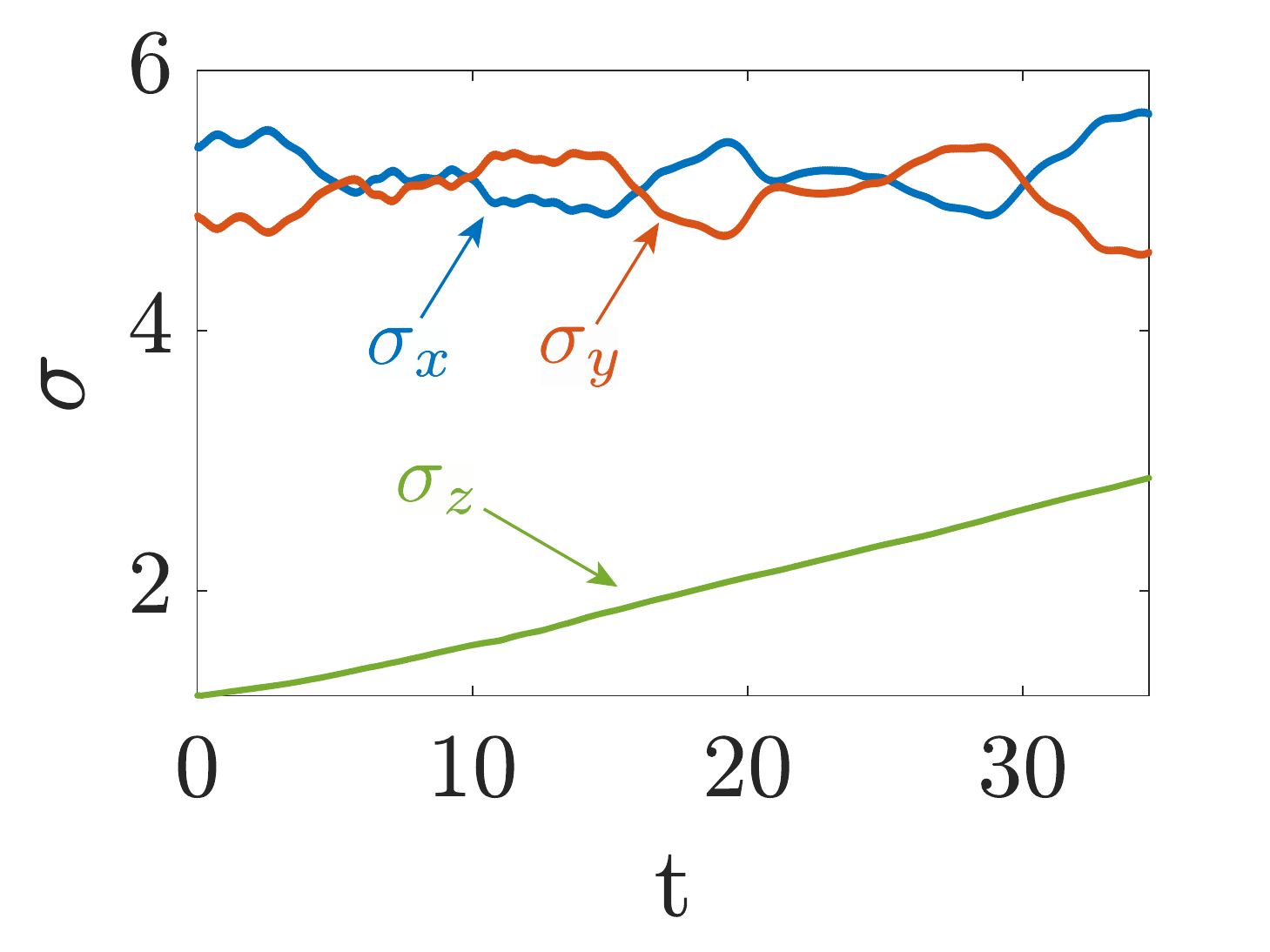}
    \includegraphics[width=0.49\linewidth]{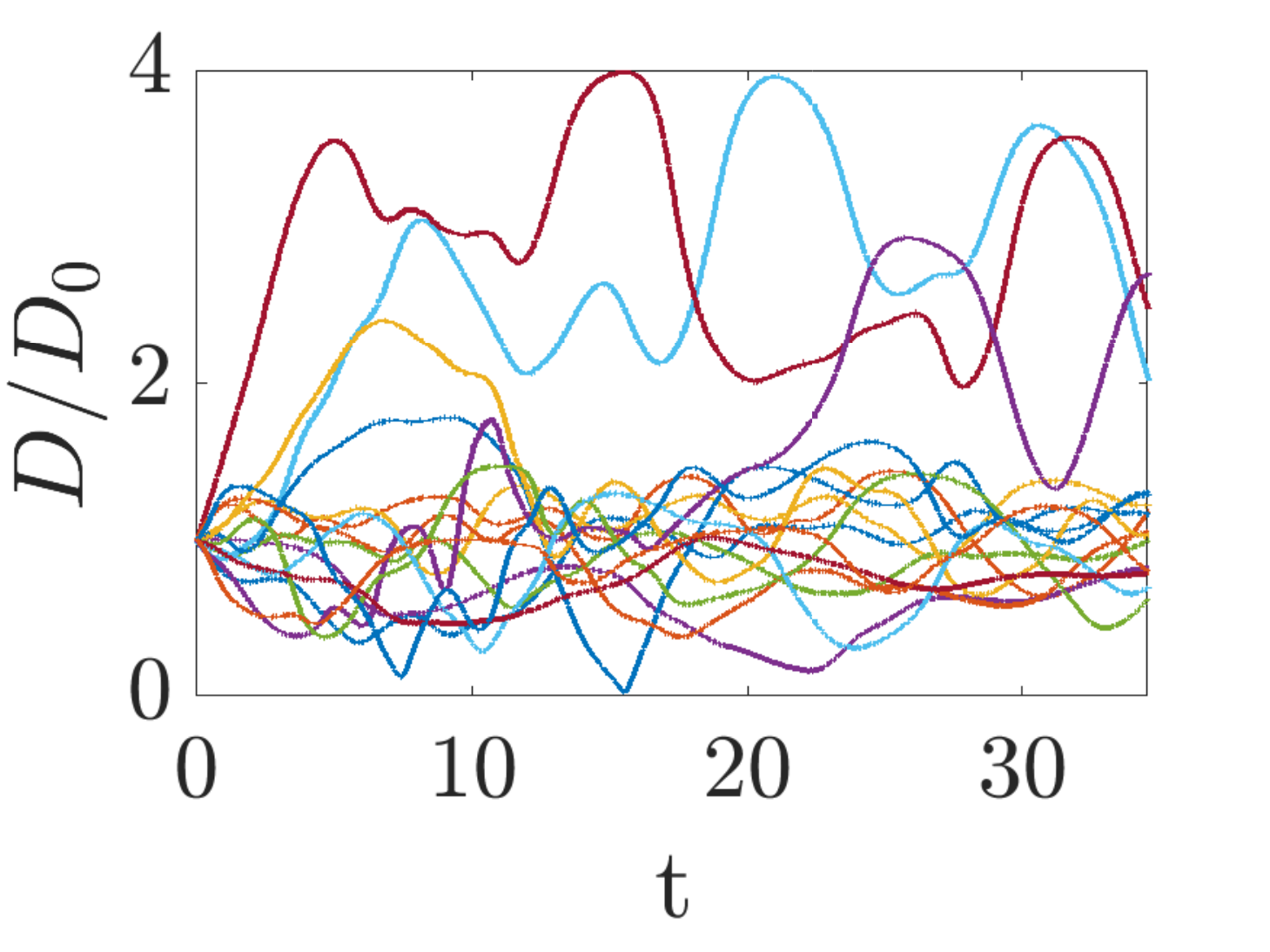}\\
    (d) \hspace{1.4in} (e)\\
    \caption{Simulation results for 16 interacting swimmers of differing magnetic moments initially placed in a plane orthogonal to the magnetic field rotation direction. (a) Direction of magnetic moment vectors with respect to the direction-parametrized propulsion speed previously shown in Fig. \ref{fig:MomentDirection}. (b),(c) give trajectories of the bodies in the $xy$- and $xz$-planes respectively.  The black line indicates the path of the mean position of the swimmers. (d) Standard deviation of the position coordinates from the mean position coordinate. (e) Distance of each swimmer from the mean position as a ratio of the initial distance from the mean. }
    \label{fig:16SwimNU}
\end{figure}

In Fig. \ref{fig:16SwimNU}, simulation results for a similar arrangement of swimmers are shown, but with less uniformity.  
While in the first case, the swimmers all possessed identical magnetic moments, for this simulation the swimmers' magnetic dipole moments are nonidentical. 
The magnetic moments are randomly (Gaussian) distributed. 
The polar coordinates of these unit dipole moment vectors are shown by the red $\times$-markers in Fig. \ref{fig:16SwimNU} (a) overlaid on the plot of propulsion velocity so that they all lie within a region of similar dynamics, but with differing propulsion velocities.  
This case of magnetic swimmers with non-uniform magnetic moments is often encountered in practice, as magnetic spheres do not necessarily have identical magnetic moments.  
Furthermore in practice the initial locations of the swimmers are unlikely to be uniformly distributed. 
Therefore the initial locations of each swimmer is randomly selected (with a Gaussian distribution with a standard deviation of one spherical diameter), with the mean located at the respective uniform grid points.

With the introduction of this nonuniformity, significant differences from the ideal case of identical swimmers with a uniform initial spatial distribution are seen.  
Firstly, the nonidentical magnetic moments produce  slower propulsion speeds for the group. 
Rather, the swimmers still rotate in circular trajectories in the $xy$-plane, but with many swimmers moving inwards and outwards towards the center of the group  as shown in Figs. \ref{fig:16SwimNU} (c) and (e).  
The standard deviations $\sigma_x$ and $\sigma_y$ show small variations but when $\sigma_x$ increases, $\sigma_y$ decreases indicating that the area occupied by the group of swimmers in the $xy$-plane is nearly the same. 
The distance ratio $D/D_0$ for each swimmer shows larger variations, but this ratio is bounded for each swimmer showing the coherence of the group of swimmers. 
In Fig. \ref{fig:16SwimNU} we see that the standard deviation in the $z$-coordinates  increases steadily, contrary to the ideal case of identical swimmers.  
This is due to differing magnetic dipole moments that produce differing propulsion speeds among the group of swimmers leading to increased separation in the $z$-direction.  

\begin{figure}
    \centering
    \includegraphics[width=0.49\linewidth]{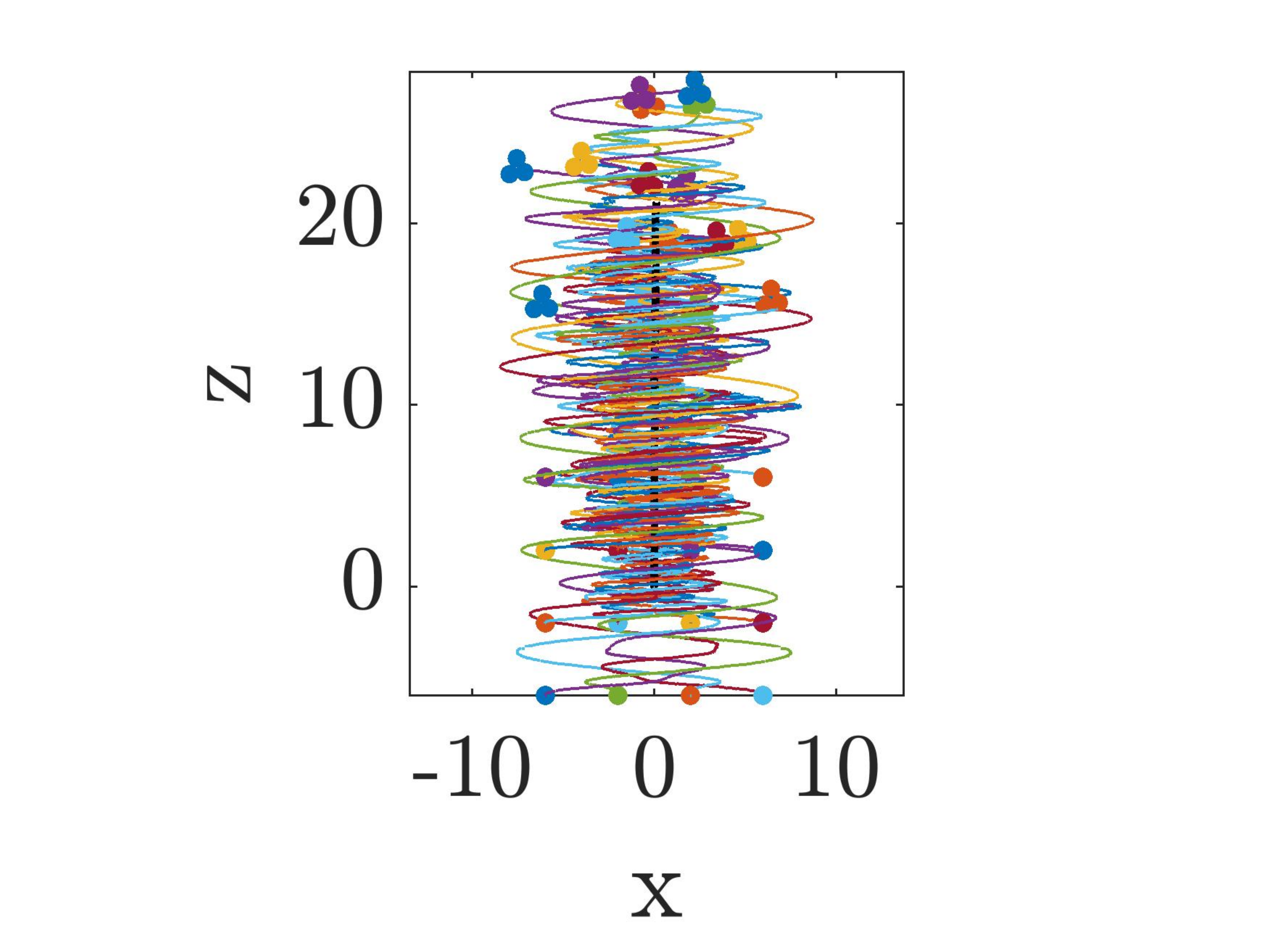}
    \includegraphics[width=0.49\linewidth]{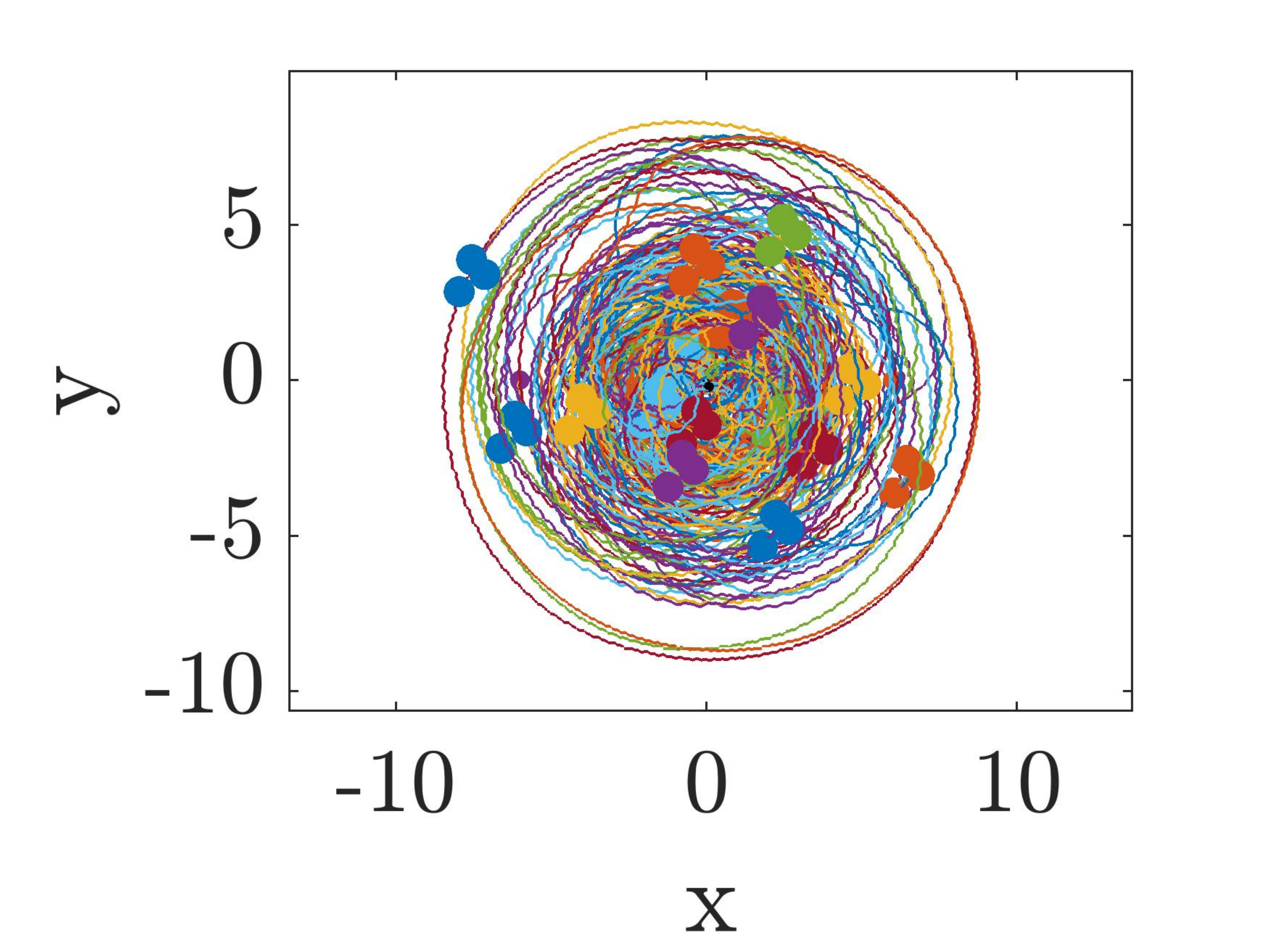}\\
    (a) \hspace{1.4in} (b)\\
    \includegraphics[width=0.49\linewidth]{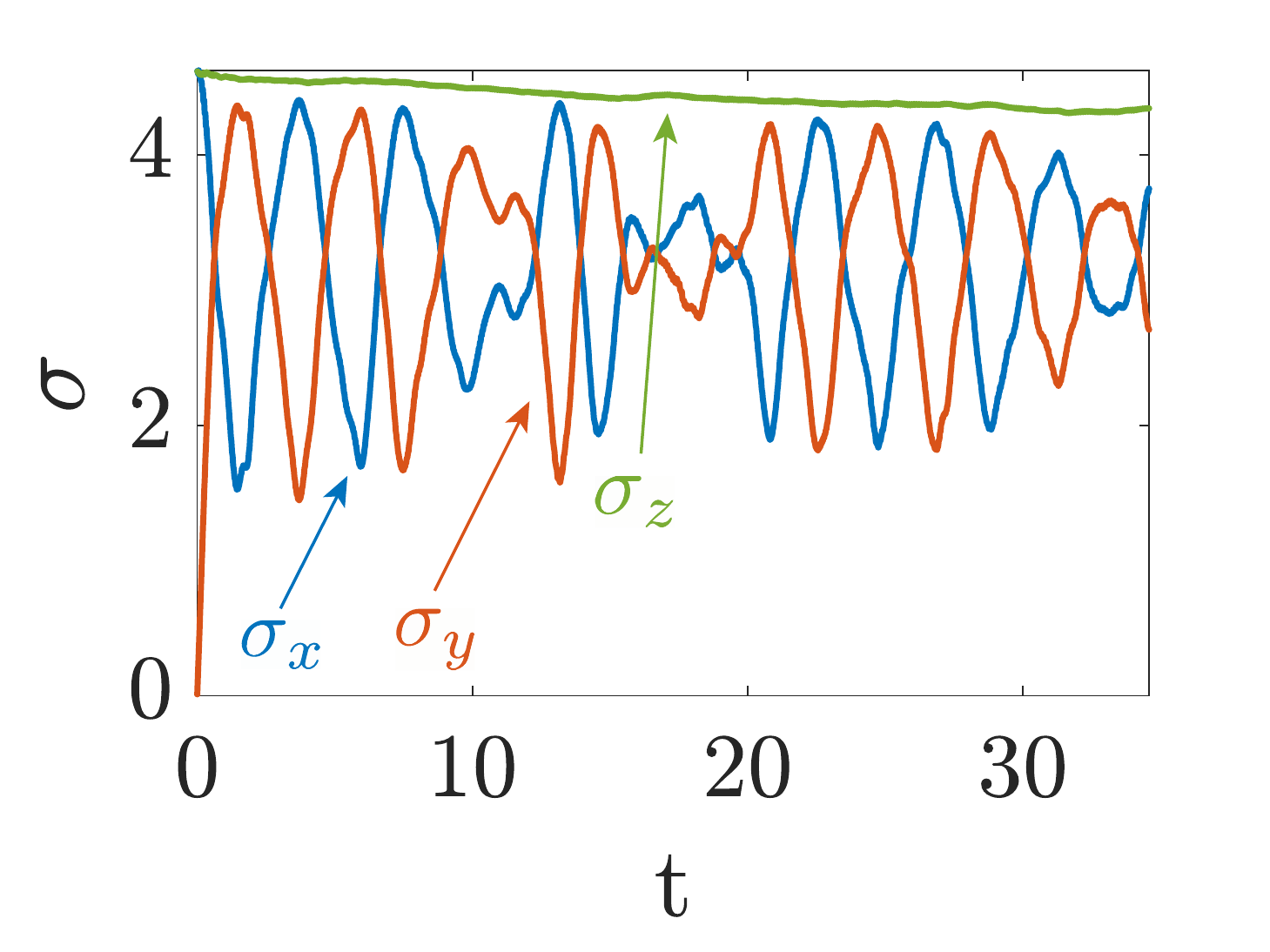}
    \includegraphics[width=0.49\linewidth]{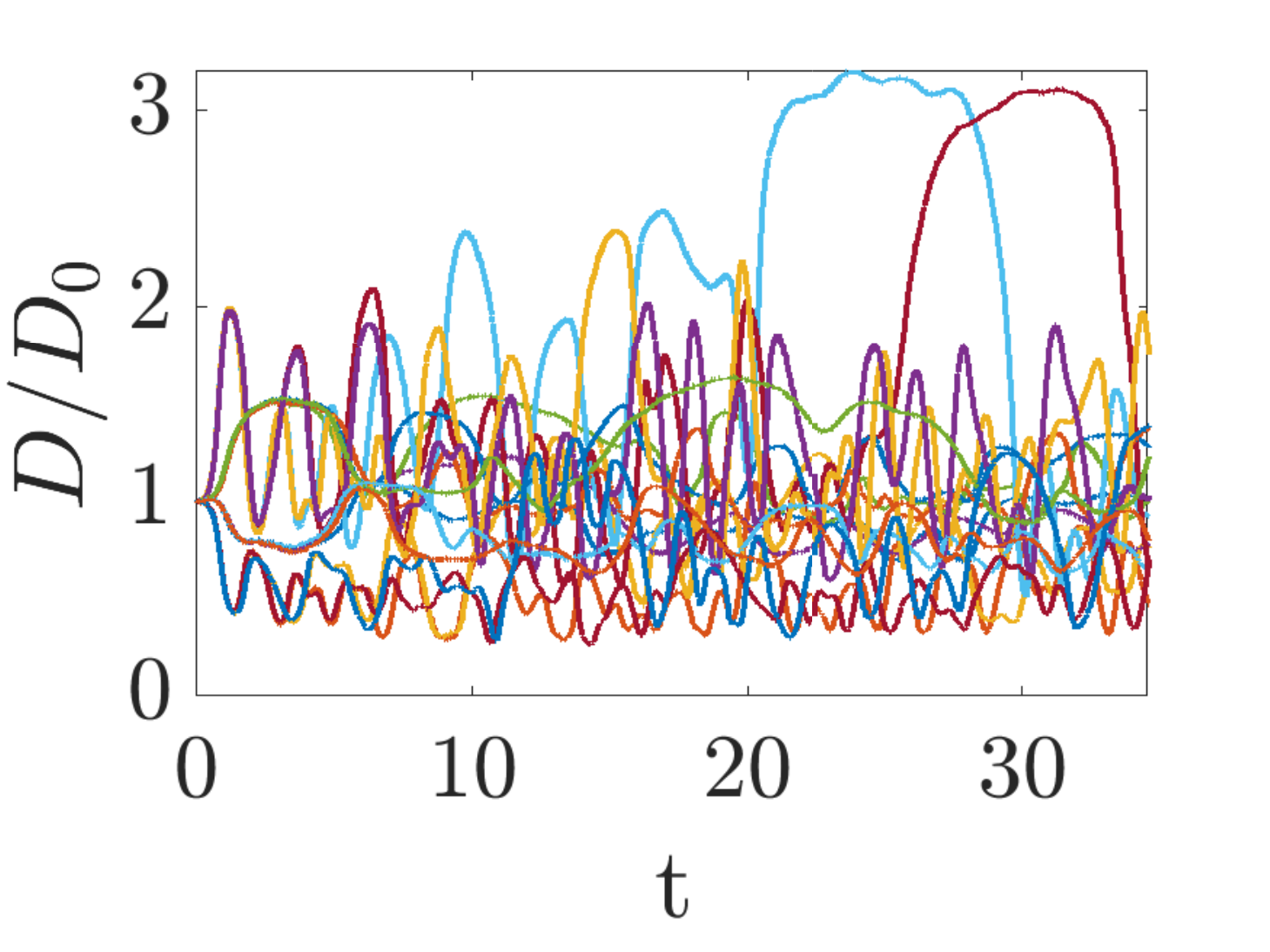}
    (c) \hspace{1.4in} (d)\\
    \caption{ Simulation results for 16 identical interacting swimmers initially placed in a plane containing the magnetic field rotation direction.  (a),(b) give trajectories of the bodies in the $xy$- and $xz$- planes.  The black line indicates the path of the mean position of the swimmers. Filled circles indicate the starting position, while the swimmer bodies indicate the final position and orientation. (c) Standard deviation of the position coordinates from the mean position coordinate. (d) Distance of each swimmer from the mean position as a ratio of the initial distance from the mean.}
    \label{fig:16SwimTandem}
\end{figure}

As with the two swimmer case,  the motion of a group of microswimmers when the rotation axis of the magnetic field lies in the plane that initially contains the group of sixteen swimmers is now considered. 
Fig. \ref{fig:16SwimTandem} gives the results of this simulation. 
Again, the magnetic field rotates in the $z$-direction by Eq. \eqref{eq:MagField} and the swimmers are placed on a uniform grid with initial separations of 4 spherical diameters, but this time in the $xz$-plane. 
In this case, it appears that swimmers that start at the same height in the $z$-direction tend to interact as a group and travel together in a spiraling trajectory, but with little interaction with swimmers starting higher or lower in the grid. 
This aligns with the results from the previous cases in which swimmers that lied in the same plane orthogonal to the direction of magnetic field rotation tend to interact, with very little hydrodynamic interaction among swimmers lying in different planes orthogonal to the magnetic field.  
Studying the standard deviations of the swimmers from the mean position in Fig. \ref{fig:16SwimTandem} (c), again it is seen that there is little separation in the $z$ direction throughout the simulation, and little net growth in the $x$- or $y$- directions either, as they oscillate between larger and smaller values as the swimmers complete helical trajectories in space.
In Fig. \ref{fig:16SwimTandem}(d), the plot of swimmer distance ratios $D/D_0$  indicates that the lateral displacements of most swimmers from the axis of net motion is small. 
Some swimmers show large motion in such lateral direction, but while some swimmers move temporarily away from the axis of net motion, others move towards this axis, but such that the area enclosed by the swimmers in the $xy$-plane remains nearly constant and undergoes little distortion.  
\begin{figure}
    \centering
    \includegraphics[width=0.55\linewidth]{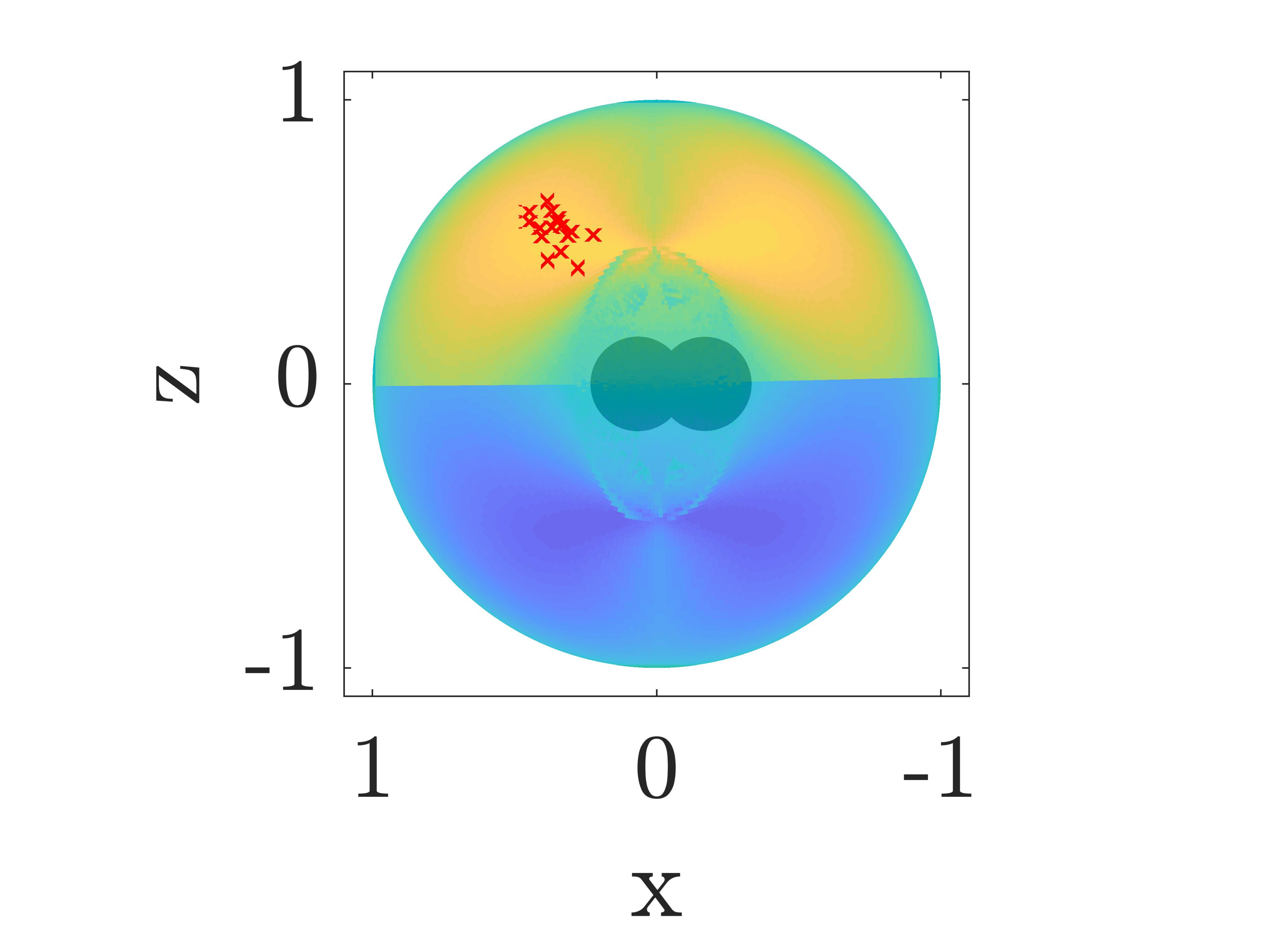}\\
    (a)\\
    \includegraphics[width=0.49\linewidth]{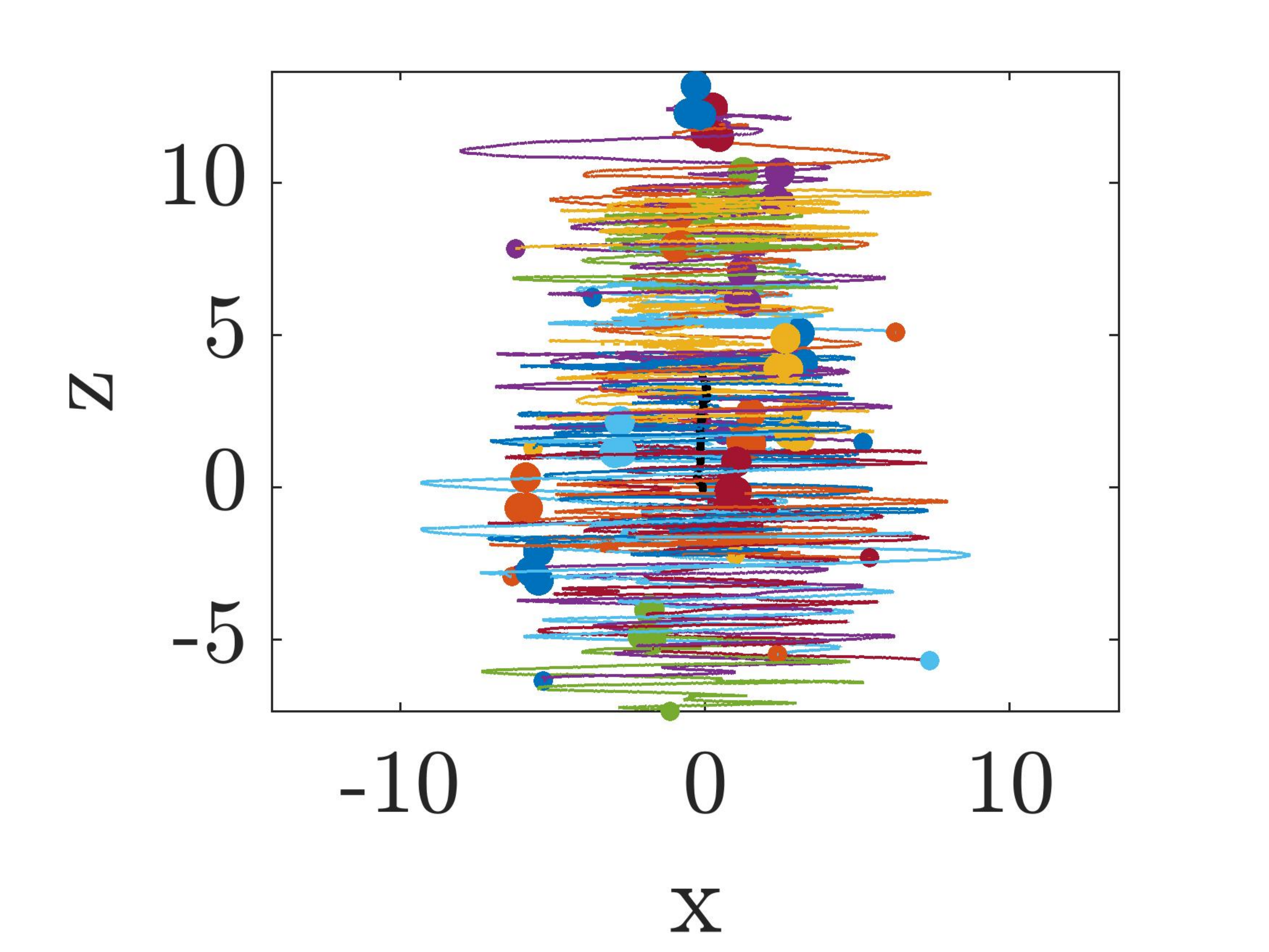}
    \includegraphics[width=0.49\linewidth]{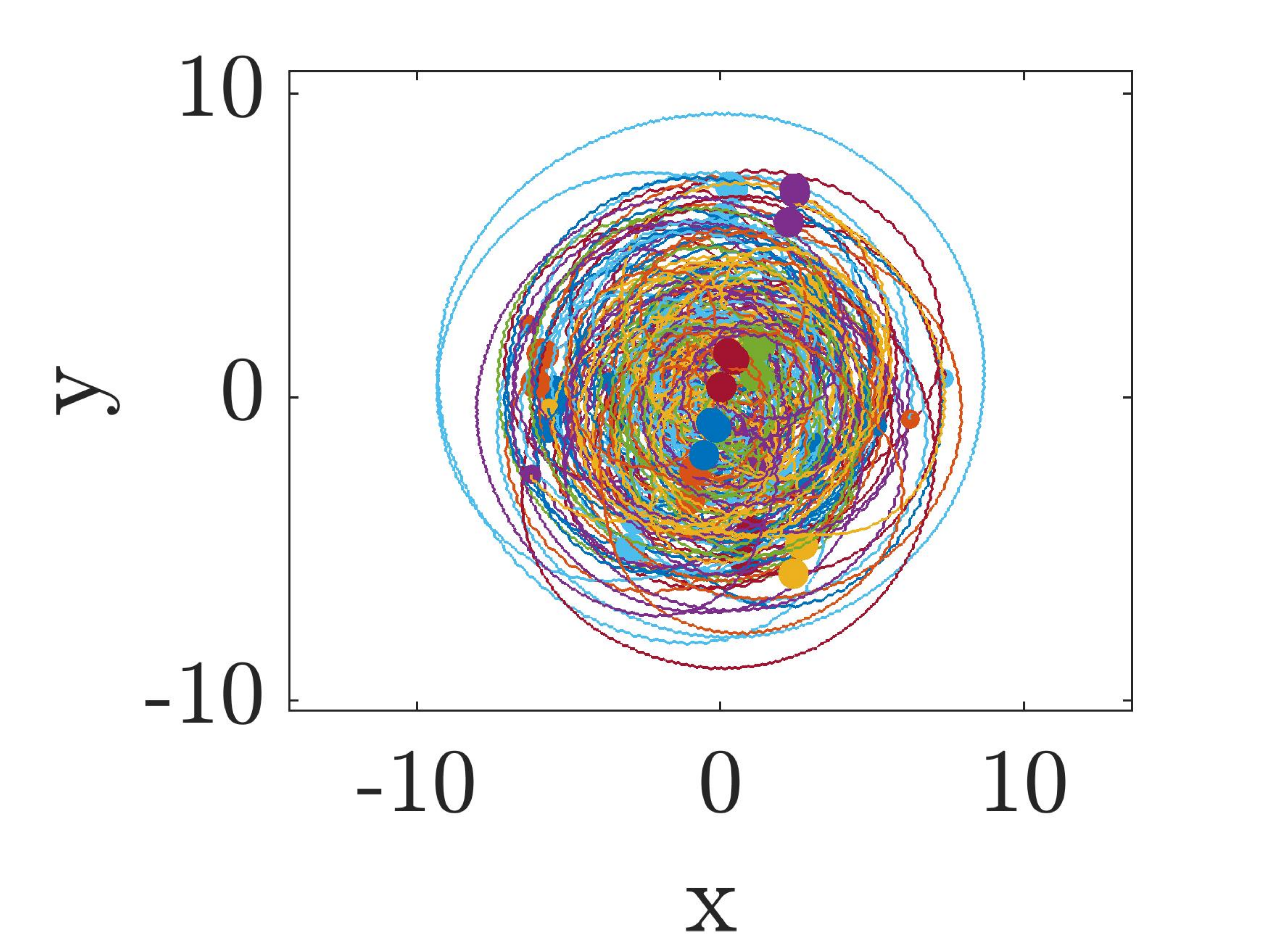}\\
    (b) \hspace{1.4in} (c)\\
    \includegraphics[width=0.49\linewidth]{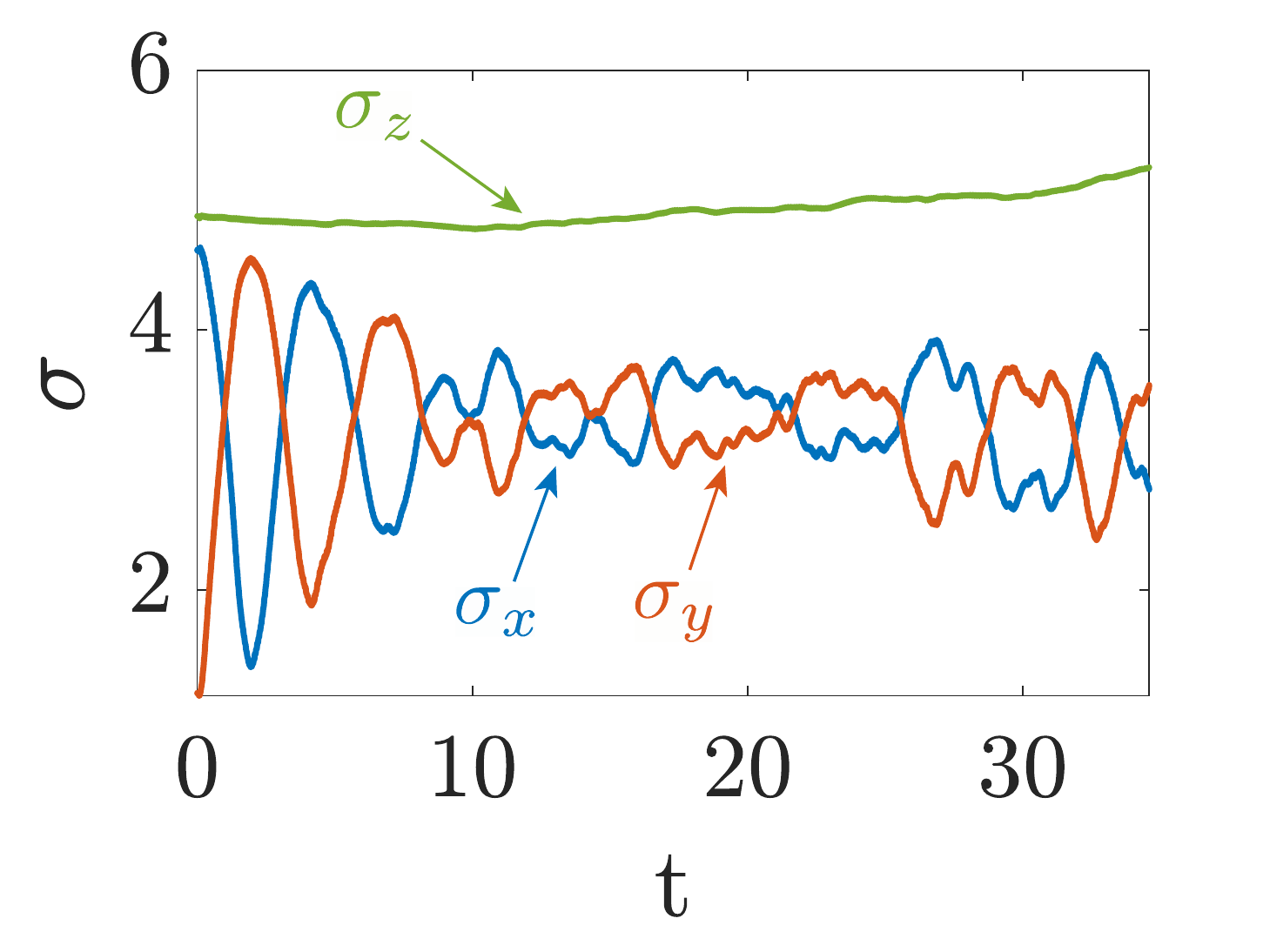}
    \includegraphics[width=0.49\linewidth]{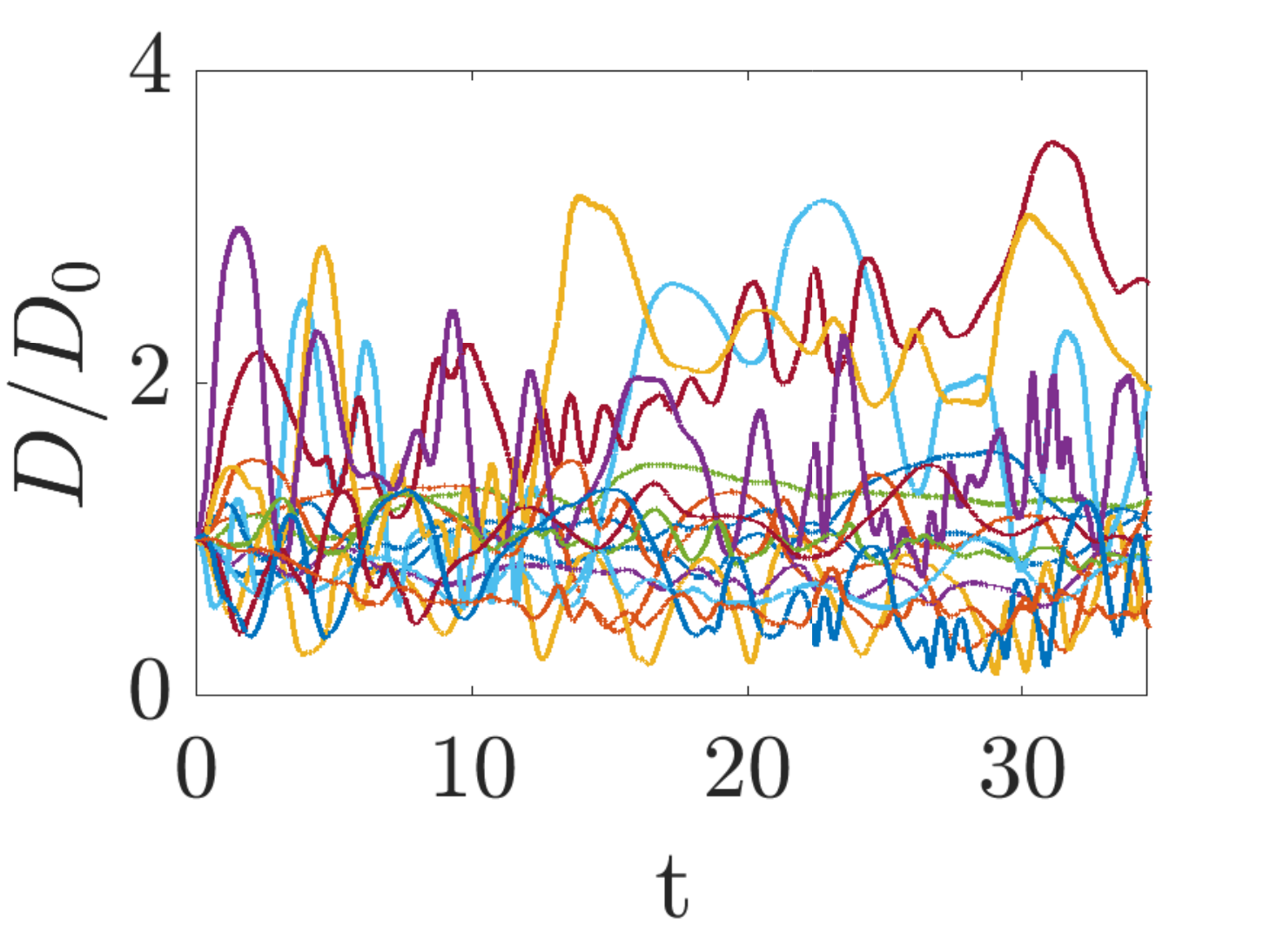}\\
    (d) \hspace{1.4in} (e)\\
    \caption{ Simulation results for 16 interacting swimmers initially placed in a plane containing the magnetic field rotation direction.  (a) direction of  magnetic  moment  vectors  with  respect  to  the  direction-parametrized  propulsion  speed  previously  shown  in  Fig.   2. (b),(c) give trajectories of the bodies in the $xy$- and $xz$- planes.  The black line indicates the path of the mean position of the swimmers. Filled circles indicate the starting position, while the swimmer bodies indicate the final position and orientation. (d) Standard deviation of the position coordinates from the mean position coordinate. (e) Distance of each swimmer from the mean position as a ratio of the initial distance from the mean.}
    \label{fig:16SwimTandemNU}
\end{figure}

In Fig. \ref{fig:16SwimTandemNU}, a non-uniform version of the case considered in Fig. \ref{fig:16SwimTandem} is given.  Here, the swimmer locations are initialized along a square grid with in-line separation distances of 4 spherical diameters, but with an added random variation to each coordinate of zero mean and standard deviation of one spherical diameter.  
As in the case of the simulation in Fig. \ref{fig:16SwimNU}, the magnetic moment orientations are also randomized as shown in Fig. \ref{fig:16SwimTandemNU} (a). 
In the plot of $D/D_0$ shown in Fig. \ref{fig:16SwimTandemNU} (e), it is shown that the group maintains its coherence, as this ratio remains near unity for most of the swimmers in the group.
This is further confirmed by studying the standard deviations of the swimmer positions from the mean position, which shows that there is not significant growth in the deviation in any of the coordinates, as was seen in the previous case that included such nonuniformity. 

\section{Conclusion}
In this paper we have presented results of the effects of hydrodynamic interactions on groups of magnetically-driven artificial microswimmers.  
Through Stokesian dynamics simulations, we analyzed the fluid velocity field generated by the motion of a single swimmer, showing that it produces a highly rotational fluid velocity field in the plane orthogonal to the magnetic field rotation and a much more rapidly decaying velocity field in planes containing this axis.  
We show that the rotating fluid velocity field produced is almost the same as that of a rotlet, the point-torque singularity of Stokes flows.
That is, when this velocity field is time-averaged over the period of magnetic field rotations, it is almost the same as that of a rotlet with a spin oriented in the direction of net locomotion.  
We then apply this finding to develop an understanding of pairwise swimmer motion, focusing on two configurations.  
We see that while little hydrodynamic interaction takes place when the magnetic field rotation is directed along the line joining the swimmers, there is significant interaction when the rotation is orthogonal to the plane containing the swimmers. 
This is explained in terms of the fluid velocity field produced by the swimmer and the approximation of the swimmer as a rotlet. 
In both cases the distance of each swimmer from the group center of mobility remains nearly constant. 
The group center itself moves like a magnetically propelled swimmer. 
We then extend this simulation to study the motion of a larger group of sixteen of these swimmers in different group configurations.  
We show that these swimmers still tend to move as a coherent group by studying the standard deviation of the swimmers' positions from the group from the group center position, which remain nearly constant. 
Further, we see that in the case of uniform magnetizations, the group center of mobility tends to remain constant in the plane orthogonal to the magnetic field rotation.  
We then introduce non-uniformity in the magnetizations and initial positions of the swimmers in the group and show that this group coherence and conservation of the mean position in the plane perpendicular to rotations is still largely maintained for this more realistic simulation. 
The results given here will be useful in developing simplified dynamical systems models and control strategies for cooperative motion of small groups of these artificial microswimmers based on spatial ensemble statistics of the group, a task which will be necessary if this developing technology is to reach its potential in a range of applications, such as targeted drug delivery. 

\begin{acknowledgments}
The authors would like to acknowledge Clemson University for the generous allotment of computation time and resources on the Palmetto Cluster. 
\end{acknowledgments}

\bibliographystyle{unsrt}
\bibliography{magswimmer}

\end{document}